\documentclass[10pt,twocolumn,reprint,amsmath,amssymb,aps,prl,showpacs,longbibliography,superscriptaddress]{revtex4-1}
\usepackage[latin9]{inputenc}
\usepackage{amsmath}
\usepackage{amssymb}
\usepackage{graphicx}

\makeatletter
\usepackage{amsfonts}
\usepackage{graphicx}
\usepackage{graphics}

\makeatother

\begin{document}
\title{Exact quasiparticle properties of a heavy polaron in BCS Fermi superfluids}
\author{Jia Wang}
\affiliation{Centre for Quantum Technology Theory, Swinburne University of Technology,
Melbourne 3122, Australia}
\author{Xia-Ji Liu}
\affiliation{Centre for Quantum Technology Theory, Swinburne University of Technology,
Melbourne 3122, Australia}
\author{Hui Hu}
\affiliation{Centre for Quantum Technology Theory, Swinburne University of Technology,
Melbourne 3122, Australia}
\date{\today}
\begin{abstract}
We present the Ramsey response and radio-frequency spectroscopy of
a heavy impurity immersed in an interacting Fermi superfluid, using
exact functional determinant approach. We describe the Fermi superfluid
through the conventional Bardeen-Cooper-Schrieffer theory and investigate
the role of the pairing gap on quasiparticle properties revealed by
the two spectroscopies. The energy cost for pair breaking prevents
Anderson\textquoteright s orthogonality catastrophe that occurs in
a non-interacting Fermi gas and allows the existence of polaron quasiparticles
in the exactly solvable heavy impurity limit. Hence, we rigorously
confirm the remarkable features such as dark continuum, molecule-hole
continuum and repulsive polaron. For a magnetic impurity scattering
at finite temperature, we predict additional resonances related to
the sub-gap Yu-Shiba-Rusinov bound state, whose positions can be used
to measure the superfluid pairing gap. For a non-magnetic scattering
at zero temperature, we surprisingly find undamped repulsive polarons.
These exact results might be readily observed in quantum gas experiments
with Bose-Fermi mixtures that have a large-mass ratio.
\end{abstract}
\maketitle
Thanks to the unprecedented controllability recently achieved in ultracold
quantum gases, investigations on non-equilibrium quantum dynamics
in many-body systems have progressed rapidly \citep{Bloch2008RMP}.
One of such intriguing problems is how a quantum gas medium responds
to a suddenly introduced impurity \citep{Bruun2014Review,Schmidt2018Review}.
The quantum gas can be either a degenerate Fermi gas or a Bose-Einstein
condensate (BEC). The impurity-medium interaction can essentially
be tuned arbitrarily via Feshbach resonance \citep{Chin2010RMP},
and a variety of impurities, such as Rydberg atoms \citep{Pfau2013Nature,Jia2015PRL,Sous2020PRResearh}
or quantum rotor \citep{Schmidt2015PRL}, can be introduced. A unique
advantage of these impurity-medium systems is that they present probably
the simplest non-trivial many-particle problem, where the medium response
can be directly measured (i.e., by Ramsey and radio-frequency spectroscopies)
and efficiently calculated even in the non-perturbative strong-coupling
regime \citep{Schmidt2018Review,Demler2012PRX}. Consequently, they
can serve as a critical meeting point for theoretical and experimental
efforts to understand the complicated quantum dynamics of interacting
many-particle systems.

Historically, the first research of impurity-medium systems in 1933
led Landau to introduce a general concept of polarons -- quasiparticles
formed by dressing the impurity with elementary excitations of the
medium \citep{Landau1933PhysZSoviet}. The new platform of ultracold
quantum gases has enabled the exploration of polaron quasiparticle
properties in a controllable and quantitative manner over the last
decade, both experimentally \citep{Schirotzek2009PRL,Zhang2012PRL,Grimm2012Nature,Kohl2012Nature,Demler2016Science,Hu2016PRL,Jorgensen2016PRL,Scazza2017PRL,Yan2019PRL,Zwierlein2020Science,Sagi2020PRX}
and theoretically \citep{Chevy2006PRA,Lobo2006PRL,Combescot2007PRL,Punk2009PRA,Cui2010PRA,Mathy2011PRL,Schmidt2012PRA,Rath2013PRA,Shashi2014PRA,Li2014PRA,Kroiss2015PRL,Levinsen2015PRL,HuHui2016PRA,Goulko2016PRA,HuHui2018PRA,PenaArdila2019PRA,Mulkerin2019AnnPhys,Jia2019PRL,Isaule2021PRA,Pessoa2021PRA,Seetharam2021PRL}.
In particular, a number of salient features of polarons have been
predicted by approximate theories and Monte Carlo simulations, including
the excited repulsive polaron with finite lifetime \citep{Cui2010PRA}
and the dark continuum \citep{Goulko2016PRA} and molecule-hole continuum
\citep{Bruun2014Review} that separate the attractive and repulsive
polaron branches. While the repulsive polaron has been unambiguously
observed in experiments \citep{Grimm2012Nature,Kohl2012Nature}, the
existence of the dark and molecule-hole continua remains elusive due
to the uncertainty in theoretical calculations. The purpose of this
Letter is to present an \emph{exact} calculation of polaron quasiparticle
properties in the heavy impurity limit and in the experimentally unexplored
regime with a Fermi superfluid medium \citep{Nishida2015PRL,Yi2015PRA,Pierce2019PRL,HuHui2021arXiv,Bigue2022arXiv}. 

Our work naturally extends the well-known exactly solvable many-body
problem of the Fermi-edge singularity of x-ray absorption spectra
in metals \citep{Nozieres1969PR,Mahan2000Book}, which is the first
and most important example of non-equilibrium many-body physics \citep{Weiss1999Book,Rosch1999AdvPhys}.
In this impurity-medium problem, the suddenly introduced infinitely
heavy impurity can excite particle-hole pairs close to Fermi surfaces
without costing finite recoil energy \citep{Schmidt2018Review,Demler2012PRX}.
The multiple particle-hole excitations completely changes the many-particle
states in the limit of a large particle number. As a result, the many-particle
states with and without impurity become orthogonal, i.e., Anderson's
``orthogonality catastrophe'' (OC) \citep{Anderson1967PRL}. In
the context of ultracold quantum gases, the Fermi-edge singularity
has been quantitatively re-examined via the functional determinant
approach (FDA) \citep{Leonid1996JMathPhys,Klich2003Book,Schonhammer2007PRB,Ivanov2013JMathPhys},
providing insightful understanding of polaron physics \citep{Bruun2014Review,Schmidt2018Review,Demler2012PRX}.
Unfortunately, strictly speaking, due to OC the attractive and repulsive
polarons do not exist, as indicated by the vanishing quasiparticle
residue \citep{Schmidt2018Review,Demler2012PRX}.

\begin{figure}
\includegraphics[width=1\columnwidth]{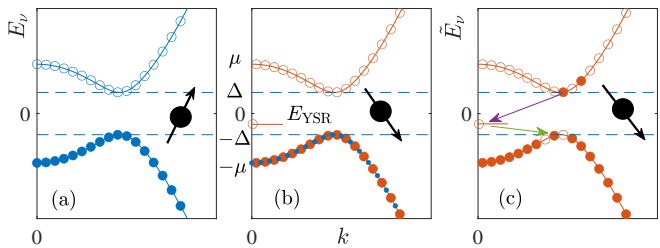}\caption{A sketch of the occupation and structure of the single-particle dispersion
spectrum of a two-component superfluid Fermi gas with a positive chemical
potential $\mu>0$ and the presence of a static impurity (black dot).
(a) shows the spectrum $E_{\nu}$ when the impurity is in the non-interacting
state (black arrow up) at zero temperature. When the impurity is in
the interacting polaron state (black arrow down), the spectrum $\tilde{E}_{\nu}$
are shown in (b) at zero and (c) finite temperature. \label{fig:Sketch}}
\end{figure}

Here, we propose an exactly solvable model of a heavy impurity immersed
in a Fermi superfluid medium described by the standard Bardeen-Cooper-Schrieffer
(BCS) pairing theory \citep{Leggett1980JPhys,Nozieres1985JLowTempPhys,HuHui2006EurphysLett}.
As multiple particle-hole excitations can be efficiently suppressed
by the energy cost of pairing breaking, Anderson's OC is avoided and
polarons acquire \emph{nonzero} quasiparticle residue. Therefore,
we obtain a benchmark theoretical model with well-defined polaron
quasiparticles, in which all the speculated characteristics of polarons
can be rigorously examined. Our results are also highly experimentally
relevant, as a BCS Fermi superfluid (of $^{6}$Li or $^{40}$K atoms)
has now been routinely realized using Feshbach resonance at the so-called
BEC-BCS crossover and a heavy atomic species such as $^{133}$Cs can
be manipulated at will as impurity. 

In Fig. \ref{fig:Sketch}, we outline an injection scheme of interest
in this work: the impurity is driven from a non-interacting hyperfine
state {[}Fig. \ref{fig:Sketch} (a){]} into an interacting state {[}Fig.
\ref{fig:Sketch} (b){]} at time $t=0$. The dynamical evolution at
later time, namely the Ramsey response, is then exactly calculated
via an extension of the FDA, from which we extract the spectral function.
The existence of pairing gap prevents the OC and preserves well-defined
polaron quasiparticle features in the spectral function. In addition
to rigorously confirming the remarkable characteristics of polarons,
our exact results also reveal two novel unique features related to
the Fermi superfluid medium: the resonances related to the sub-gap
Yu-Shiba-Rusinov (YSR) state bound to a magnetic impurity \citep{Yu1965ActaPhysSin,Shiba1968ProgTheorPhys,Rusinov1969JETP,Vernier2011PRA,Jiang2011PRA}
and repulsive polarons with infinitely long lifetime in the case of
a non-magnetic impurity scattering.

\textbf{\textit{Theory}}\textsl{.} The fundamental Ramsey response
is the real-time overlap function between the many-body state with
and without the impurity, $S(t)=\langle\mathrm{e}^{\mathrm{i}\hat{\mathcal{H}}_{i}t}\mathrm{e}^{-\mathrm{i}\hat{\mathcal{H}}_{f}t}\rangle\equiv{\rm Tr}[\mathrm{e}^{\mathrm{i}\hat{\mathcal{H}}_{i}t}\mathrm{e}^{-\mathrm{i}\hat{\mathcal{H}}_{f}t}\hat{\rho}_{\mathrm{0}}]$,
where $\mathcal{H}_{i}$ ($\mathcal{H}_{f}$) is the many-body Hamiltonian
in the absence (presence) of the impurity scattering and $\hat{\rho}_{\mathrm{0}}$
is the initial state of the Fermi system (Hereafter, we use the units
of $\hbar\equiv1$). Complementarily, the frequency-resolved spectral
function $A(\omega)={\rm Re}\int_{0}^{\infty}e^{i\omega t}S(t)dt/\pi$,
which determines the radio-frequency (rf) spectroscopy, can be obtained
by a Fourier transformation \citep{Goold2011PRA,Demler2012PRX}. Since
the complexity of the many-body Hamiltonians increases exponentially
with the numbers of particles $N$ in the system, an exact calculation
of $S(t)$ is usually inaccessible. However, in the case that $\mathcal{H}_{i}$
and $\mathcal{H}_{f}$ are both fermionic, bilinear many-body operators,
the overlap function can reduce to a determinant in single-particle
Hilbert space that grows only linearly to $N$ \citep{Leonid1996JMathPhys,Klich2003Book,Schonhammer2007PRB,Ivanov2013JMathPhys}:
\begin{equation}
S(t)=e^{-i\omega_{0}t}{\rm det}[1-\hat{n}+e^{i\hat{h}_{i}t}e^{-i\hat{h}_{f}t}\hat{n}],\label{eq:FDA_BCS}
\end{equation}
where $\hat{n}$ is the occupation number operator, and $\hat{h}_{i}$
($\hat{h}_{f}$) are the single-particle representatives of $\mathcal{H}_{i}$
($\mathcal{H}_{f}$) up to some constant terms that $\omega_{0}$
compensates. For example, $\mathcal{H}_{f}=K_{0}+\omega_{0}+\int d\mathbf{r}\hat{\phi}^{\dagger}(\mathbf{r})\underline{h_{f}(\mathbf{r})}\hat{\phi}(\mathbf{r})$,
where $K_{0}$ is an unimportant constant and 
\begin{equation}
\underline{h_{f}(\mathbf{r})}=\left(\begin{array}{cc}
-\frac{\nabla^{2}}{2m}+V_{\uparrow}(\mathbf{r})-\mu & \Delta\\
\Delta & \frac{\nabla^{2}}{2m}-V_{\downarrow}(\mathbf{r})+\mu
\end{array}\right),\label{eq:hf}
\end{equation}
with $V_{\sigma}(\mathbf{r})$ being the potential between impurity
and $\sigma$-component fermion. Note that, here we already extend
the FDA to the case of a BCS Fermi superfluid, which is characterized
by the pairing gap $\Delta$ and chemical potential $\mu$ to be determined
by a given scattering length $a$ between unlike fermions, temperature
$T$ and Fermi momentum $k_{F}=(3\pi^{2}N/\mathcal{V})^{1/3}$, where
$\mathcal{V}$ is the system volume. It is convenient to use the Nambu
spinor operators as $\hat{\phi}^{\dagger}(\mathbf{r})=[c_{\uparrow}^{\dagger}(\mathbf{r}),c_{\downarrow}(\mathbf{r})]$,
where $c_{\sigma}^{\dagger}(\mathbf{r})$ {[}$c_{\sigma}(\mathbf{r})${]}
being the creation (annihilation) operator for a $\sigma$-component
fermion at position $\mathbf{r}$. We also have $\omega_{0}={\rm Tr}\underline{V_{\downarrow}}$,
which corresponds to the phase factor in Eq. (\ref{eq:FDA_BCS}) with
$\underline{V_{\downarrow}}$ being the matrix format of $V_{\downarrow}(\mathbf{r})$
in a complete orthogonal set of basis. Finally, $\underline{h_{i}(\mathbf{r})}$
can be obtained by setting $V_{\sigma}(\mathbf{r})$ equals zero in
Eq. (\ref{eq:hf}). In what follows, we briefly describe the computation
procedure and present our main physical results, but relegate numerical
details and additional discussions to a complementary paper \citep{AccompanyingLong2022PRA}.

We consider a finite system confined in a sphere of radius $R$ and
take the system size towards infinity, while keeping the density constant,
until numerical results are converged \citep{Demler2012PRX}. We focus
on the $s$-wave channel and use finite-range potentials $V_{\sigma}(r)$
whose corresponding energy-dependent scattering length $a_{\sigma}(E_{F})=-\tan\eta_{\sigma}(k_{F})/k_{F}$,
where $\eta_{\sigma}(k_{F})$ is the $s$-wave scattering length between
the impurity and $\sigma$-component fermions at the Fermi energy
$E_{F}=k_{F}^{2}/(2m)$ . We find that numerical results are not sensitive
to other short-range details of the potential. Therefore, for simplicity
we denote $a_{\sigma}\equiv a_{\sigma}(E_{F})$ hereafter \citep{ScatteringLength}.
Finally, for a given set of parameters $\{k_{F},a,a_{\uparrow},a_{\downarrow},T\}$,
we can calculate Eq. (\ref{eq:FDA_BCS}) by finding the eigenpairs
$E_{\nu}$, $\phi_{\nu}\equiv[\phi_{\nu,\uparrow}(r),\phi_{\nu,\downarrow}(r)]$
for $\underline{h_{i}(r)}$ and $\tilde{E}_{\nu}$, $\tilde{\phi}_{\nu}$
for $\underline{h_{f}(r)}$. In this presentation, the occupation
operator $\hat{n}$ is given by a diagonal matrix with elements $n_{\nu\nu}=[e^{-E_{\nu}/(k_{B}T)}+1]^{-1}$,
where $k_{B}$ is the Boltzmann constant. Figure \ref{fig:Sketch}(a)
gives a sketch of the occupation and structure of single-particle
spectrum $E_{\nu}$ without impurity at $T=0$, which includes a completely
filled Fermi sea (filled circles in the lower branch) and an empty
one (empty circles on the top) separated by $2\Delta$. The presence
of impurity scattering leads to a shift of the single-particle levels
$E_{\nu}\rightarrow\tilde{E}_{\nu}$ as shown in Fig. \ref{fig:Sketch}(b),
where the small blue dots shows $E_{\nu}$ for comparison. In the
case of a magnetic impurity scattering ($a_{\uparrow}\ne a_{\downarrow}$),
there also exists a sub-gap YSR bound state with energy $E_{{\rm YSR}}$
\citep{Yu1965ActaPhysSin,Shiba1968ProgTheorPhys,Rusinov1969JETP,Vernier2011PRA}.

\begin{figure}
\includegraphics[width=1\columnwidth]{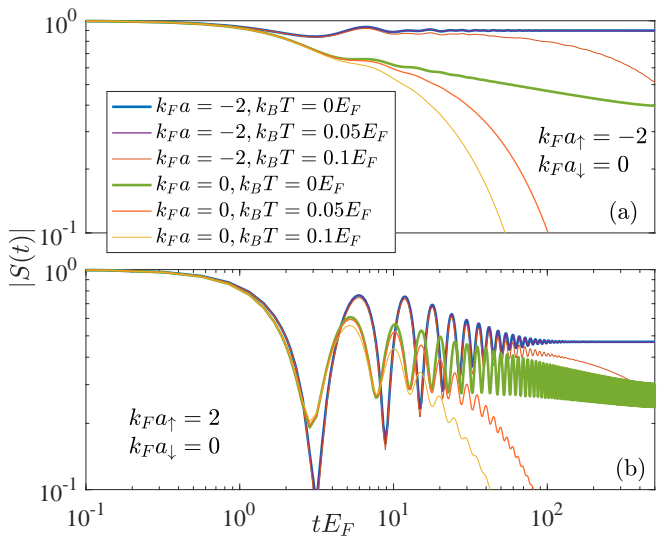}

\caption{Ramsey response $|S(t)|$ for a magnetic impurity scattering with
(a) attractive scattering lengths $a_{\uparrow}<0$ and (b) repulsive
scattering lengths $a_{\uparrow}>0$ are shown for different values
of the scattering length $a$ between the two-component fermions and
different temperature $T$; see legend. \label{fig:St}}
\end{figure}

\textbf{\textit{Ramsey response}}\textsl{.} As reported in Fig. \ref{fig:St},
our numerical examples here focus on the BCS side of the crossover
$k_{F}a=-2<0$, where $\mu\approx0.85E_{F}$ and $\Delta\approx0.40E_{F}$
at zero temperature. While our method applies to the whole crossover
regime, mean-field description becomes only qualitatively reliable
on the BEC side. We also focus on the simplest case, where the impurity
only interacts with the spin-up component fermion, i.e., $V_{\downarrow}(r)=0$.
For comparison, we show also the results for $k_{F}a=0$ (with $\mu=E_{F}$
and $\Delta=0$), where the $\uparrow$-component of medium reduces
to a non-interacting Fermi gas that couples with the impurity, and
the $\downarrow$-component being simply a spectator. These results
agree with previous studies for both zero and finite temperature \citep{Demler2012PRX}. 

At $k_{F}a=0$ and $T=0$, the asymptotic behavior of $|S(t)|$ at
large $t$ exhibits a power-law decay $|S(t)|\sim t^{-\alpha}$, reflecting
Anderson\textquoteright s OC and x-ray infrared singularity $\epsilon^{\alpha-1}$
at the low-energy scale set by the inverse time $\epsilon\sim\hbar/t$
\citep{Nozieres1969PR,Schmidt2018Review,Demler2012PRX}. In contrast,
in the presence of a pairing gap $\Delta\neq0$, $|S(t)|\sim t^{0}$
at $T=0$, indicating OC is prevented as the low-energy scale is now
cut by $\Delta$ \citep{Ma1985PRB}. At finite temperature, although
both with or without the pairing gap, $|S(t)|$ shows an exponential
decay at large $t$, such behavior appears at a much later time for
finite $\Delta$. In particular, for nonzero $\Delta$ the results
at $T=0$ and $T=0.05E_{F}/k_{B}$ are almost overlapping at $tE_{F}\le500$,
showing that the pairing gap can also protect the response signal
against thermal fluctuation if $k_{B}T\ll\Delta$. 

More quantitatively, at nonzero pairing gap we have the following
analytic result,

\begin{equation}
S(t\rightarrow\infty)\simeq\begin{cases}
D_{a}e^{-iE_{a}t}, & a_{\uparrow}<0\\
D_{a}e^{-iE_{a}t}+D_{r}e^{-iE_{r}t}, & a_{\uparrow}>0
\end{cases}\label{eq:StLarge}
\end{equation}
which fits excellently well to our numerical results, with $D_{a}$
$E_{a}$, $D_{r}$, and $E_{r}$ being fitting parameters. At small
$\Delta$, the coefficients $|D_{a}|\propto(\Delta/E_{F})^{\alpha_{a}}$
and $|D_{r}|\propto(\Delta/E_{F})^{\alpha_{r}}$, making the asymptotic
form that agrees with the modification of the analytic expression
of $S(t)$ for a non-interacting Fermi gas medium via replacing the
low-energy cut-off $1/t\rightarrow\Delta$ (see, i.e., Eqs. (12) and
(15) of Ref. \citep{Demler2012PRX}). However, our numerical results
indicate the power-law exponents $\alpha$ and $\alpha_{a}$ are close
to but not exactly the same as the analytical results given in \citep{Demler2012PRX},
see Ref. \citep{AccompanyingLong2022PRA} for details. At $T=0$,
$E_{a}$ is purely real and corresponds to the attractive polaron
energy that satisfies $E_{a}=\sum_{\nu}n_{\nu\nu}(E_{\nu}-\tilde{E}_{\nu})$,
indicating that the attractive polaron can be regarded as the renormalization
of the Fermi sea due to the impurity level. The repulsive polaron
energy $E_{r}$ is in general complex, where we denote the real and
imaginary part as ${\rm Re}E_{r}$ and ${\rm Im}E_{r}$. 

\begin{figure}[t]
\includegraphics[width=1\columnwidth]{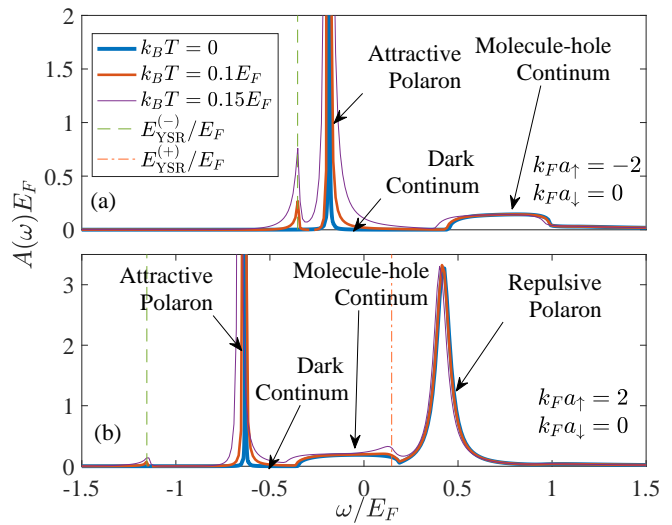}\caption{The spectral function $A(\omega)$ at $k_{F}a=-2$ for different temperature
(see legend) and different scattering length $a_{\uparrow}<0$ in
(a) and $a_{\uparrow}>0$ in (b). \label{fig:Aw2D}}
\end{figure}

\begin{figure*}
\includegraphics[width=1\textwidth]{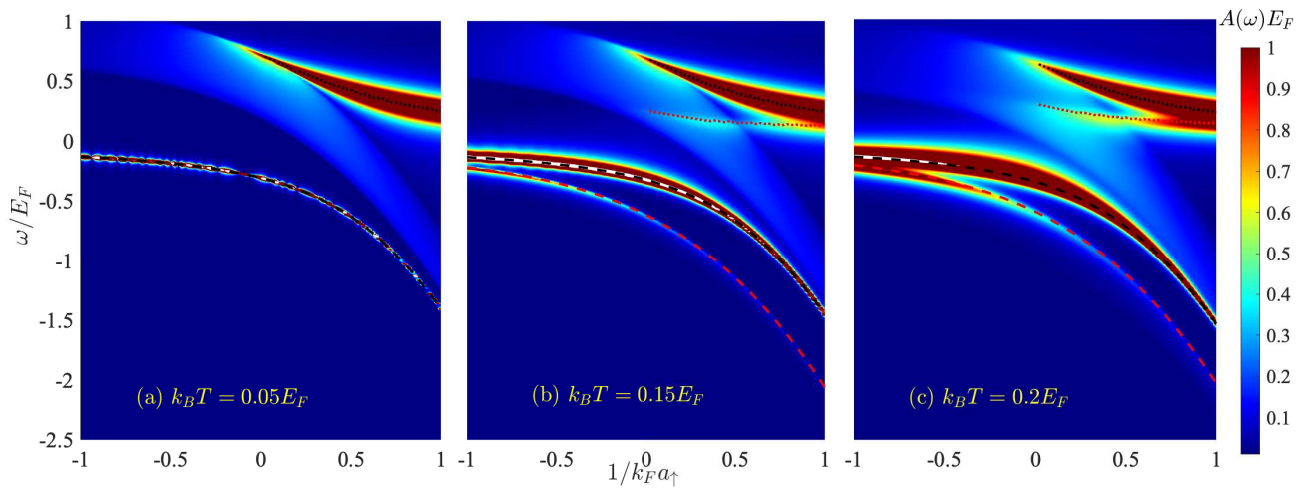}\caption{The spectral function $A(\omega)$ at $k_{F}a=-2$ as a function of
$a_{\uparrow}$ for different temperature: $k_{B}T=0.05,\ 0.15,\ 0.2E_{F}$
for (a), (b) and (c), respectively. The black dashed (dotted) curves
show the attractive (repulsive) polaron energy and the red dashed
(dotted) curves show $E_{{\rm YSR}}^{(+)}$ ($E_{{\rm YSR}}^{(-)}$)
at finite temperature. \label{fig:Aw3D}}
\end{figure*}

\textbf{\textit{rf-spectroscopy}}. One of the key observations of
this Letter is the saturation of $\left|S(t)\right|$ at large time,
which implies a finite polaron quasiparticle residue $Z=\left|D_{a}\right|\propto\Delta^{\alpha_{a}}$.
To check this, we calculate the frequency response $A(\omega)$ accurately
with a Fourier transformation of $S(t)$. We choose a large cut-off
$t^{*}\sim500/E_{F}$, evaluate $S(t)$ numerically for $t<t^{*}$
and use the fitting formula in Eq. (\ref{eq:StLarge}) for $t\ge t^{*}$.
As shown in Fig. \ref{fig:Aw2D} by thick blue solid curves for zero-temperature
results, the attractive polaron is characterized by a $\delta$-function
peak at $E_{a}$ (with a small artificial width for visibility), unambiguously
confirming the existence of a well-defined quasiparticle. The attractive
polaron peak separates with a molecule-hole continuum by a region
of anomalously low spectral weight, namely the ``dark continuum''.
This spectral gap has previously been shown in other polaron systems
with approximate calculations, where the anomalously low spectral
weight might be an artifact of the adopted approximations. Only recently,
a diagrammatic Monte Carlo study indicates that this dark continuum
might be indeed physical \citep{Goulko2016PRA}. Here, the heavy polaron
spectral function is calculated via FDA, and hence can be regarded
as an exact proof of the dark continuum. For $a>0$, a Lorentzian
lineshape with a peak at ${\rm Re}(E_{r})$ corresponds to the repulsive
polaron. The finite width determined by ${\rm Im}(E_{r})$ implies
that the repulsive polaron has a finite lifetime.

In Fig. \ref{fig:Aw2D}, finite-temperature results are indicated
by the red thin (purple thinner) curves for $k_{B}T=0.1E_{F}\ (0.15E_{F})$.
Other than the expected thermal broadening, some additional surprising
features show up. An enhancement of spectral weight appears sharply
at the energy $E_{{\rm YSR}}^{(-)}=E_{a}-(\Delta-E_{{\rm YSR}})$
below the attractive polaron. This spectral feature corresponds to
the decay process highlighted by the purple arrow in Fig. \ref{fig:Sketch}(c),
where an additional particle initially excited to the upper Fermi
sea by thermal fluctuation is driven to the the YSR state. For the
case of $k_{F}a_{\uparrow}>0$, a feature associated with the repulsive
polaron appears at $E_{{\rm YSR}}^{(+)}={\rm Re}(E_{r})-(E_{{\rm YSR}}+\Delta)$,
as indicated by the green arrow in Fig. \ref{fig:Sketch}(c): an additional
particle decays from the YSR state to the lower Fermi sea. These features
can be better observed in the whole spectrum of $a_{\uparrow}$ across
a resonance, as shown in Fig. \ref{fig:Aw3D} for different temperatures.
The YSR features are negligible at $k_{B}T=0.05E_{F}$, and the spectrum
in Fig. \ref{fig:Aw3D}(a) is almost the same as zero-temperature
results. This shows the protection against finite temperature provided
by the pairing gap. The YSR features become apparent in Fig. \ref{fig:Aw3D}(b)
at $k_{B}T=0.15E_{F}$ and shows broadening at $k_{B}T=0.2E_{F}$.
We emphasize that this range of temperature is accessible for current
experiments. The polaron spectrum can be applied to measure the superfluid
gap $\Delta$ and $E_{{\rm YSR}}$. In particular, we notice, on the
positive side $a_{\uparrow}>0$, if $E_{a}$, ${\rm Re}(E_{r})$,
$E_{{\rm YSR}}^{(-)}$ and $E_{{\rm YSR}}^{(+)}$ can all be measured
accurately, we have $2\Delta=E_{a}+{\rm Re}(E_{r})-E_{{\rm YSR}}^{(-)}-E_{{\rm YSR}}^{(+)}$
that does not depend on $E_{{\rm YSR}}$. Since this formula only
relies on the existence of the gap and a mid-gap state, we anticipate
it can be used to measure $\Delta$ accurately for a Fermi superfluid
that can not be quantitatively described by the BCS theory.

Finally, we discuss briefly our observations for the case of a non-magnetic
impurity scattering with $a_{\uparrow}=a_{\downarrow}$, where the
YSR features are absent as expected. Interestingly, we also discover
that the repulsive polaron exhibits itself as a $\delta$-function
peak in the spectral function at zero temperature. We believe the
underlying physics might be due to the gapless density fluctuations
in the Fermi superfluid excited by the perfect balance of the two
scattering lengths. As a result, the impurity couples to phonon excitations
of the superfluid and forms a long-lived repulsive polaron. For more
details, we refer to the complementary paper \citep{AccompanyingLong2022PRA}.

\textbf{\textit{Experimental realization}}\textsl{.} Our predictions
can be readily confirmed by immersing heavy $^{133}$Cs impurities
in a BCS Fermi superfluid of $^{6}$Li atoms routinely observed near
a broad Feshbach resonance $B_{0}\simeq832$ G. The two interspecies
broad resonances located nearby at $843$ G and $889$ G \citep{Tung2013PRA}
allow us to independently control the $^{133}$Cs-$^{6}$Li scattering
lengths $a_{\uparrow,\downarrow}$. Both magnetic and non-magnetic
impurity scatterings can therefore be realized by tuning the magnetic
field \citep{AccompanyingLong2022PRA}.

\textbf{\textit{Conclusions}}\textsl{.} We have calculated the response
functions of driving a heavy impurity in a BCS superfluid from non-interacting
to interacting hyperfine states. Due to the existence of a pairing
gap in the superfluid, the OC is prevented and genuine polaron quasiparticles
exit. The underlying physical reason is apparent: exciting particle-hole
pairs in this system requires energy cost for Cooper-pair breaking,
and hence multiple particle-hole excitations are energetic unfavored.
We emphasize that our FDA can support this conclusion since it is
essentially exact, unlike some approximations such as extended Chevy\textquoteright s
ansatz \citep{Chevy2006PRA,Cui2010PRA} or $T$-matrix method \citep{Bruun2014Review,HuHui2018PRA}
that allow only a few particle-hole excitations. In this respect,
our calculation can be regarded as an exact theoretical model of polarons.
Many features of the spectrum structure, such as the existence of
a $\delta$-function peak for the attractive polaron and a dark continuum,
are rigorously confirmed to be universal. The pairing gap also protects
the polaron against thermal fluctuation, preserving clear polaron
features in response functions at a finite temperature $k_{B}T\sim\Delta$.
Furthermore, we discover that the polaron spectrum can be applied
to measure the background superfluid excitation spectrum, such as
the pairing gap $\Delta$. Interestingly, in the case of a magnetic
impurity, the polaron spectrum at finite but low temperature has sharp
features that can be used to measure the sub-gap YSR bound state.
For non-magnetic impurity, we predict the existence of a long-lived
repulsive polaron.
\begin{acknowledgments}
We are grateful to Xing-Can Yao for insightful discussions. This research
was supported by the Australian Research Council's (ARC) Discovery
Program, Grants No. DE180100592 and No. DP190100815 (J.W.), and Grant
No. DP180102018 (X.-J.L). 
\end{acknowledgments}

\bibliography{RefHeavyCrossoverPolaron}

\begin{thebibliography}{69}%
\makeatletter
\providecommand \@ifxundefined [1]{%
 \@ifx{#1\undefined}
}%
\providecommand \@ifnum [1]{%
 \ifnum #1\expandafter \@firstoftwo
 \else \expandafter \@secondoftwo
 \fi
}%
\providecommand \@ifx [1]{%
 \ifx #1\expandafter \@firstoftwo
 \else \expandafter \@secondoftwo
 \fi
}%
\providecommand \natexlab [1]{#1}%
\providecommand \enquote  [1]{``#1''}%
\providecommand \bibnamefont  [1]{#1}%
\providecommand \bibfnamefont [1]{#1}%
\providecommand \citenamefont [1]{#1}%
\providecommand \href@noop [0]{\@secondoftwo}%
\providecommand \href [0]{\begingroup \@sanitize@url \@href}%
\providecommand \@href[1]{\@@startlink{#1}\@@href}%
\providecommand \@@href[1]{\endgroup#1\@@endlink}%
\providecommand \@sanitize@url [0]{\catcode `\\12\catcode `\$12\catcode
  `\&12\catcode `\#12\catcode `\^12\catcode `\_12\catcode `\%12\relax}%
\providecommand \@@startlink[1]{}%
\providecommand \@@endlink[0]{}%
\providecommand \url  [0]{\begingroup\@sanitize@url \@url }%
\providecommand \@url [1]{\endgroup\@href {#1}{\urlprefix }}%
\providecommand \urlprefix  [0]{URL }%
\providecommand \Eprint [0]{\href }%
\providecommand \doibase [0]{http://dx.doi.org/}%
\providecommand \selectlanguage [0]{\@gobble}%
\providecommand \bibinfo  [0]{\@secondoftwo}%
\providecommand \bibfield  [0]{\@secondoftwo}%
\providecommand \translation [1]{[#1]}%
\providecommand \BibitemOpen [0]{}%
\providecommand \bibitemStop [0]{}%
\providecommand \bibitemNoStop [0]{.\EOS\space}%
\providecommand \EOS [0]{\spacefactor3000\relax}%
\providecommand \BibitemShut  [1]{\csname bibitem#1\endcsname}%
\let\auto@bib@innerbib\@empty
%</preamble>
\bibitem [{\citenamefont {Bloch}\ \emph {et~al.}(2008)\citenamefont {Bloch},
  \citenamefont {Dalibard},\ and\ \citenamefont {Zwerger}}]{Bloch2008RMP}%
  \BibitemOpen
  \bibfield  {author} {\bibinfo {author} {\bibfnamefont {Immanuel}\
  \bibnamefont {Bloch}}, \bibinfo {author} {\bibfnamefont {Jean}\ \bibnamefont
  {Dalibard}}, \ and\ \bibinfo {author} {\bibfnamefont {Wilhelm}\ \bibnamefont
  {Zwerger}},\ }\bibfield  {title} {\enquote {\bibinfo {title} {Many-body
  physics with ultracold gases},}\ }\href@noop {} {\bibfield  {journal}
  {\bibinfo  {journal} {Rev. Mod. Phys.}\ }\textbf {\bibinfo {volume} {80}},\
  \bibinfo {pages} {885--964} (\bibinfo {year} {2008})}\BibitemShut {NoStop}%
\bibitem [{\citenamefont {Massignan}\ \emph {et~al.}(2014)\citenamefont
  {Massignan}, \citenamefont {Zaccanti},\ and\ \citenamefont
  {Bruun}}]{Bruun2014Review}%
  \BibitemOpen
  \bibfield  {author} {\bibinfo {author} {\bibfnamefont {P.}~\bibnamefont
  {Massignan}}, \bibinfo {author} {\bibfnamefont {M.}~\bibnamefont {Zaccanti}},
  \ and\ \bibinfo {author} {\bibfnamefont {G.~M.}\ \bibnamefont {Bruun}},\
  }\bibfield  {title} {\enquote {\bibinfo {title} {Polarons, dressed molecules
  and itinerant ferromagnetism in ultra-cold {Fermi} gases},}\ }\href@noop {}
  {\bibfield  {journal} {\bibinfo  {journal} {Rep. Prog. Phys.}\ }\textbf
  {\bibinfo {volume} {77}},\ \bibinfo {pages} {034401} (\bibinfo {year}
  {2014})}\BibitemShut {NoStop}%
\bibitem [{\citenamefont {Schmidt}\ \emph {et~al.}(2018)\citenamefont
  {Schmidt}, \citenamefont {Knap}, \citenamefont {Ivanov}, \citenamefont {You},
  \citenamefont {Cetina},\ and\ \citenamefont {Demler}}]{Schmidt2018Review}%
  \BibitemOpen
  \bibfield  {author} {\bibinfo {author} {\bibfnamefont {R.}~\bibnamefont
  {Schmidt}}, \bibinfo {author} {\bibfnamefont {M.}~\bibnamefont {Knap}},
  \bibinfo {author} {\bibfnamefont {D.~A.}\ \bibnamefont {Ivanov}}, \bibinfo
  {author} {\bibfnamefont {J.-S.}\ \bibnamefont {You}}, \bibinfo {author}
  {\bibfnamefont {M.}~\bibnamefont {Cetina}}, \ and\ \bibinfo {author}
  {\bibfnamefont {E.}~\bibnamefont {Demler}},\ }\bibfield  {title} {\enquote
  {\bibinfo {title} {Universal many-body response of heavy impurities coupled
  to a {Fermi} sea: a review of recent progress},}\ }\href@noop {} {\bibfield
  {journal} {\bibinfo  {journal} {Rep. Prog. Phys.}\ }\textbf {\bibinfo
  {volume} {81}},\ \bibinfo {pages} {024401} (\bibinfo {year}
  {2018})}\BibitemShut {NoStop}%
\bibitem [{\citenamefont {Chin}\ \emph {et~al.}(2010)\citenamefont {Chin},
  \citenamefont {Grimm}, \citenamefont {Julienne},\ and\ \citenamefont
  {Tiesinga}}]{Chin2010RMP}%
  \BibitemOpen
  \bibfield  {author} {\bibinfo {author} {\bibfnamefont {Cheng}\ \bibnamefont
  {Chin}}, \bibinfo {author} {\bibfnamefont {Rudolf}\ \bibnamefont {Grimm}},
  \bibinfo {author} {\bibfnamefont {Paul}\ \bibnamefont {Julienne}}, \ and\
  \bibinfo {author} {\bibfnamefont {Eite}\ \bibnamefont {Tiesinga}},\
  }\bibfield  {title} {\enquote {\bibinfo {title} {Feshbach resonances in
  ultracold gases},}\ }\href@noop {} {\bibfield  {journal} {\bibinfo  {journal}
  {Rev. Mod. Phys.}\ }\textbf {\bibinfo {volume} {82}},\ \bibinfo {pages}
  {1225--1286} (\bibinfo {year} {2010})}\BibitemShut {NoStop}%
\bibitem [{\citenamefont {Balewski}\ \emph {et~al.}(2013)\citenamefont
  {Balewski}, \citenamefont {Krupp}, \citenamefont {Gaj}, \citenamefont
  {Peter}, \citenamefont {B{'"u}chler}, \citenamefont {L{\"o}w}, \citenamefont
  {Hofferberth},\ and\ \citenamefont {Pfau}}]{Pfau2013Nature}%
  \BibitemOpen
  \bibfield  {author} {\bibinfo {author} {\bibfnamefont {Jonathan~B.}\
  \bibnamefont {Balewski}}, \bibinfo {author} {\bibfnamefont {Alexander~T.}\
  \bibnamefont {Krupp}}, \bibinfo {author} {\bibfnamefont {Anita}\ \bibnamefont
  {Gaj}}, \bibinfo {author} {\bibfnamefont {David}\ \bibnamefont {Peter}},
  \bibinfo {author} {\bibfnamefont {Hans~Peter}\ \bibnamefont {B{'"u}chler}},
  \bibinfo {author} {\bibfnamefont {Robert}\ \bibnamefont {L{\"o}w}}, \bibinfo
  {author} {\bibfnamefont {Sebastian}\ \bibnamefont {Hofferberth}}, \ and\
  \bibinfo {author} {\bibfnamefont {Tilman}\ \bibnamefont {Pfau}},\ }\bibfield
  {title} {\enquote {\bibinfo {title} {Coupling a single electron to a
  bose-einstein condensate},}\ }\href@noop {} {\bibfield  {journal} {\bibinfo
  {journal} {Nature (London)}\ }\textbf {\bibinfo {volume} {502}},\ \bibinfo
  {pages} {664--667} (\bibinfo {year} {2013})}\BibitemShut {NoStop}%
\bibitem [{\citenamefont {Wang}\ \emph {et~al.}(2015)\citenamefont {Wang},
  \citenamefont {Gacesa},\ and\ \citenamefont {C{\^o}t{\'e}}}]{Jia2015PRL}%
  \BibitemOpen
  \bibfield  {author} {\bibinfo {author} {\bibfnamefont {Jia}\ \bibnamefont
  {Wang}}, \bibinfo {author} {\bibfnamefont {Marko}\ \bibnamefont {Gacesa}}, \
  and\ \bibinfo {author} {\bibfnamefont {R.}~\bibnamefont {C{\^o}t{\'e}}},\
  }\bibfield  {title} {\enquote {\bibinfo {title} {Rydberg electrons in a
  {Bose-Einstein} condensate},}\ }\href@noop {} {\bibfield  {journal} {\bibinfo
   {journal} {Phys. Rev. Lett.}\ }\textbf {\bibinfo {volume} {114}},\ \bibinfo
  {pages} {243003} (\bibinfo {year} {2015})}\BibitemShut {NoStop}%
\bibitem [{\citenamefont {Sous}\ \emph {et~al.}(2020)\citenamefont {Sous},
  \citenamefont {Sadeghpour}, \citenamefont {Killian}, \citenamefont {Demler},\
  and\ \citenamefont {Schmidt}}]{Sous2020PRResearh}%
  \BibitemOpen
  \bibfield  {author} {\bibinfo {author} {\bibfnamefont {John}\ \bibnamefont
  {Sous}}, \bibinfo {author} {\bibfnamefont {H.~R.}\ \bibnamefont
  {Sadeghpour}}, \bibinfo {author} {\bibfnamefont {T.~C.}\ \bibnamefont
  {Killian}}, \bibinfo {author} {\bibfnamefont {Eugene}\ \bibnamefont
  {Demler}}, \ and\ \bibinfo {author} {\bibfnamefont {Richard}\ \bibnamefont
  {Schmidt}},\ }\bibfield  {title} {\enquote {\bibinfo {title} {Rydberg
  impurity in a fermi gas: Quantum statistics and rotational blockade},}\
  }\href {\doibase 10.1103/PhysRevResearch.2.023021} {\bibfield  {journal}
  {\bibinfo  {journal} {Phys. Rev. Research}\ }\textbf {\bibinfo {volume}
  {2}},\ \bibinfo {pages} {023021} (\bibinfo {year} {2020})}\BibitemShut
  {NoStop}%
\bibitem [{\citenamefont {Schmidt}\ and\ \citenamefont
  {Lemeshko}(2015)}]{Schmidt2015PRL}%
  \BibitemOpen
  \bibfield  {author} {\bibinfo {author} {\bibfnamefont {Richard}\ \bibnamefont
  {Schmidt}}\ and\ \bibinfo {author} {\bibfnamefont {Mikhail}\ \bibnamefont
  {Lemeshko}},\ }\bibfield  {title} {\enquote {\bibinfo {title} {Rotation of
  quantum impurities in the presence of a many-body environment},}\ }\href@noop
  {} {\bibfield  {journal} {\bibinfo  {journal} {Phys. Rev. Lett.}\ }\textbf
  {\bibinfo {volume} {114}},\ \bibinfo {pages} {203001} (\bibinfo {year}
  {2015})}\BibitemShut {NoStop}%
\bibitem [{\citenamefont {Knap}\ \emph {et~al.}(2012)\citenamefont {Knap},
  \citenamefont {Shashi}, \citenamefont {Nishida}, \citenamefont {Imambekov},
  \citenamefont {Abanin},\ and\ \citenamefont {Demler}}]{Demler2012PRX}%
  \BibitemOpen
  \bibfield  {author} {\bibinfo {author} {\bibfnamefont {Michael}\ \bibnamefont
  {Knap}}, \bibinfo {author} {\bibfnamefont {Aditya}\ \bibnamefont {Shashi}},
  \bibinfo {author} {\bibfnamefont {Yusuke}\ \bibnamefont {Nishida}}, \bibinfo
  {author} {\bibfnamefont {Adilet}\ \bibnamefont {Imambekov}}, \bibinfo
  {author} {\bibfnamefont {Dmitry~A.}\ \bibnamefont {Abanin}}, \ and\ \bibinfo
  {author} {\bibfnamefont {Eugene}\ \bibnamefont {Demler}},\ }\bibfield
  {title} {\enquote {\bibinfo {title} {Time-dependent impurity in ultracold
  fermions: Orthogonality catastrophe and beyond},}\ }\href@noop {} {\bibfield
  {journal} {\bibinfo  {journal} {Phys. Rev. X}\ }\textbf {\bibinfo {volume}
  {2}},\ \bibinfo {pages} {041020} (\bibinfo {year} {2012})}\BibitemShut
  {NoStop}%
\bibitem [{\citenamefont {Landau}(1933)}]{Landau1933PhysZSoviet}%
  \BibitemOpen
  \bibfield  {author} {\bibinfo {author} {\bibfnamefont {L.}~\bibnamefont
  {Landau}},\ }\bibfield  {title} {\enquote {\bibinfo {title} {Uber die
  bewegung der elektronen im kristallgitter},}\ }\href@noop {} {\bibfield
  {journal} {\bibinfo  {journal} {Phys. Z. Soviet.}\ }\textbf {\bibinfo
  {volume} {3}},\ \bibinfo {pages} {664} (\bibinfo {year} {1933})}\BibitemShut
  {NoStop}%
\bibitem [{\citenamefont {Schirotzek}\ \emph {et~al.}(2009)\citenamefont
  {Schirotzek}, \citenamefont {Wu}, \citenamefont {Sommer},\ and\ \citenamefont
  {Zwierlein}}]{Schirotzek2009PRL}%
  \BibitemOpen
  \bibfield  {author} {\bibinfo {author} {\bibfnamefont {Andr{\'e}}\
  \bibnamefont {Schirotzek}}, \bibinfo {author} {\bibfnamefont {Cheng-Hsun}\
  \bibnamefont {Wu}}, \bibinfo {author} {\bibfnamefont {Ariel}\ \bibnamefont
  {Sommer}}, \ and\ \bibinfo {author} {\bibfnamefont {Martin~W.}\ \bibnamefont
  {Zwierlein}},\ }\bibfield  {title} {\enquote {\bibinfo {title} {Observation
  of {Fermi} polarons in a tunable {Fermi} liquid of ultracold atoms},}\
  }\href@noop {} {\bibfield  {journal} {\bibinfo  {journal} {Phys. Rev. Lett.}\
  }\textbf {\bibinfo {volume} {102}},\ \bibinfo {pages} {230402} (\bibinfo
  {year} {2009})}\BibitemShut {NoStop}%
\bibitem [{\citenamefont {Zhang}\ \emph {et~al.}(2012)\citenamefont {Zhang},
  \citenamefont {Ong}, \citenamefont {Arakelyan},\ and\ \citenamefont
  {Thomas}}]{Zhang2012PRL}%
  \BibitemOpen
  \bibfield  {author} {\bibinfo {author} {\bibfnamefont {Y.}~\bibnamefont
  {Zhang}}, \bibinfo {author} {\bibfnamefont {W.}~\bibnamefont {Ong}}, \bibinfo
  {author} {\bibfnamefont {I.}~\bibnamefont {Arakelyan}}, \ and\ \bibinfo
  {author} {\bibfnamefont {J.~E.}\ \bibnamefont {Thomas}},\ }\bibfield  {title}
  {\enquote {\bibinfo {title} {Polaron-to-polaron transitions in the
  radio-frequency spectrum of a quasi-two-dimensional fermi gas},}\ }\href@noop
  {} {\bibfield  {journal} {\bibinfo  {journal} {Phys. Rev. Lett.}\ }\textbf
  {\bibinfo {volume} {108}},\ \bibinfo {pages} {235302} (\bibinfo {year}
  {2012})}\BibitemShut {NoStop}%
\bibitem [{\citenamefont {Kohstall}\ \emph {et~al.}(2012)\citenamefont
  {Kohstall}, \citenamefont {Zaccanti}, \citenamefont {Jag}, \citenamefont
  {Trenkwalder}, \citenamefont {Massignan}, \citenamefont {Bruun},
  \citenamefont {Schreck},\ and\ \citenamefont {Grimm}}]{Grimm2012Nature}%
  \BibitemOpen
  \bibfield  {author} {\bibinfo {author} {\bibfnamefont {C.}~\bibnamefont
  {Kohstall}}, \bibinfo {author} {\bibfnamefont {M.}~\bibnamefont {Zaccanti}},
  \bibinfo {author} {\bibfnamefont {M.}~\bibnamefont {Jag}}, \bibinfo {author}
  {\bibfnamefont {A.}~\bibnamefont {Trenkwalder}}, \bibinfo {author}
  {\bibfnamefont {P.}~\bibnamefont {Massignan}}, \bibinfo {author}
  {\bibfnamefont {G.~M.}\ \bibnamefont {Bruun}}, \bibinfo {author}
  {\bibfnamefont {F.}~\bibnamefont {Schreck}}, \ and\ \bibinfo {author}
  {\bibfnamefont {R.}~\bibnamefont {Grimm}},\ }\bibfield  {title} {\enquote
  {\bibinfo {title} {Metastability and coherence of repulsive polarons in a
  strongly interacting {Fermi} mixture},}\ }\href@noop {} {\bibfield  {journal}
  {\bibinfo  {journal} {Nature (London)}\ }\textbf {\bibinfo {volume} {485}},\
  \bibinfo {pages} {615} (\bibinfo {year} {2012})}\BibitemShut {NoStop}%
\bibitem [{\citenamefont {Koschorreck}\ \emph {et~al.}(2012)\citenamefont
  {Koschorreck}, \citenamefont {Pertot}, \citenamefont {Vogt}, \citenamefont
  {Fr{\"o}hlich}, \citenamefont {Feld},\ and\ \citenamefont
  {K{\"o}hl}}]{Kohl2012Nature}%
  \BibitemOpen
  \bibfield  {author} {\bibinfo {author} {\bibfnamefont {Marco}\ \bibnamefont
  {Koschorreck}}, \bibinfo {author} {\bibfnamefont {Daniel}\ \bibnamefont
  {Pertot}}, \bibinfo {author} {\bibfnamefont {Enrico}\ \bibnamefont {Vogt}},
  \bibinfo {author} {\bibfnamefont {Bernd}\ \bibnamefont {Fr{\"o}hlich}},
  \bibinfo {author} {\bibfnamefont {Michael}\ \bibnamefont {Feld}}, \ and\
  \bibinfo {author} {\bibfnamefont {Michael}\ \bibnamefont {K{\"o}hl}},\
  }\bibfield  {title} {\enquote {\bibinfo {title} {Attractive and repulsive
  {Fermi} polarons in two dimensions},}\ }\href@noop {} {\bibfield  {journal}
  {\bibinfo  {journal} {Nature (London)}\ }\textbf {\bibinfo {volume} {485}},\
  \bibinfo {pages} {619} (\bibinfo {year} {2012})}\BibitemShut {NoStop}%
\bibitem [{\citenamefont {Cetina}\ \emph {et~al.}(2016)\citenamefont {Cetina},
  \citenamefont {Jag}, \citenamefont {Lous}, \citenamefont {Fritsche},
  \citenamefont {M.Walraven}, \citenamefont {Grimm}, \citenamefont {Levinsen},
  \citenamefont {Parish}, \citenamefont {Schmidt}, \citenamefont {Knap},\ and\
  \citenamefont {Demler}}]{Demler2016Science}%
  \BibitemOpen
  \bibfield  {author} {\bibinfo {author} {\bibfnamefont {M.}~\bibnamefont
  {Cetina}}, \bibinfo {author} {\bibfnamefont {M.}~\bibnamefont {Jag}},
  \bibinfo {author} {\bibfnamefont {R.~S.}\ \bibnamefont {Lous}}, \bibinfo
  {author} {\bibfnamefont {I.}~\bibnamefont {Fritsche}}, \bibinfo {author}
  {\bibfnamefont {J.~T.}\ \bibnamefont {M.Walraven}}, \bibinfo {author}
  {\bibfnamefont {R.}~\bibnamefont {Grimm}}, \bibinfo {author} {\bibfnamefont
  {J.}~\bibnamefont {Levinsen}}, \bibinfo {author} {\bibfnamefont {M.~M.}\
  \bibnamefont {Parish}}, \bibinfo {author} {\bibfnamefont {R.}~\bibnamefont
  {Schmidt}}, \bibinfo {author} {\bibfnamefont {M.}~\bibnamefont {Knap}}, \
  and\ \bibinfo {author} {\bibfnamefont {E.}~\bibnamefont {Demler}},\
  }\bibfield  {title} {\enquote {\bibinfo {title} {Ultrafast many-body
  interferometry of impurities coupled to a fermi sea},}\ }\href@noop {}
  {\bibfield  {journal} {\bibinfo  {journal} {Science}\ }\textbf {\bibinfo
  {volume} {354}},\ \bibinfo {pages} {96} (\bibinfo {year} {2016})}\BibitemShut
  {NoStop}%
\bibitem [{\citenamefont {Hu}\ \emph {et~al.}(2016{\natexlab{a}})\citenamefont
  {Hu}, \citenamefont {Van~de Graaff}, \citenamefont {Kedar}, \citenamefont
  {Corson}, \citenamefont {Cornell},\ and\ \citenamefont {Jin}}]{Hu2016PRL}%
  \BibitemOpen
  \bibfield  {author} {\bibinfo {author} {\bibfnamefont {Ming-Guang}\
  \bibnamefont {Hu}}, \bibinfo {author} {\bibfnamefont {Michael~J.}\
  \bibnamefont {Van~de Graaff}}, \bibinfo {author} {\bibfnamefont {Dhruv}\
  \bibnamefont {Kedar}}, \bibinfo {author} {\bibfnamefont {John~P.}\
  \bibnamefont {Corson}}, \bibinfo {author} {\bibfnamefont {Eric~A.}\
  \bibnamefont {Cornell}}, \ and\ \bibinfo {author} {\bibfnamefont
  {Deborah~S.}\ \bibnamefont {Jin}},\ }\bibfield  {title} {\enquote {\bibinfo
  {title} {{Bose} polarons in the strongly interacting regime},}\ }\href@noop
  {} {\bibfield  {journal} {\bibinfo  {journal} {Phys. Rev. Lett.}\ }\textbf
  {\bibinfo {volume} {117}},\ \bibinfo {pages} {055301} (\bibinfo {year}
  {2016}{\natexlab{a}})}\BibitemShut {NoStop}%
\bibitem [{\citenamefont {J{\o{}}rgensen}\ \emph {et~al.}(2016)\citenamefont
  {J{\o{}}rgensen}, \citenamefont {Wacker}, \citenamefont {Skalmstang},
  \citenamefont {Parish}, \citenamefont {Levinsen}, \citenamefont
  {Christensen}, \citenamefont {Bruun},\ and\ \citenamefont
  {Arlt}}]{Jorgensen2016PRL}%
  \BibitemOpen
  \bibfield  {author} {\bibinfo {author} {\bibfnamefont {Nils~B.}\ \bibnamefont
  {J{\o{}}rgensen}}, \bibinfo {author} {\bibfnamefont {Lars}\ \bibnamefont
  {Wacker}}, \bibinfo {author} {\bibfnamefont {Kristoffer~T.}\ \bibnamefont
  {Skalmstang}}, \bibinfo {author} {\bibfnamefont {Meera~M.}\ \bibnamefont
  {Parish}}, \bibinfo {author} {\bibfnamefont {Jesper}\ \bibnamefont
  {Levinsen}}, \bibinfo {author} {\bibfnamefont {Rasmus~S.}\ \bibnamefont
  {Christensen}}, \bibinfo {author} {\bibfnamefont {Georg~M.}\ \bibnamefont
  {Bruun}}, \ and\ \bibinfo {author} {\bibfnamefont {Jan~J.}\ \bibnamefont
  {Arlt}},\ }\bibfield  {title} {\enquote {\bibinfo {title} {Observation of
  attractive and repulsive polarons in a {Bose-Einstein} condensate},}\
  }\href@noop {} {\bibfield  {journal} {\bibinfo  {journal} {Phys. Rev. Lett.}\
  }\textbf {\bibinfo {volume} {117}},\ \bibinfo {pages} {055302} (\bibinfo
  {year} {2016})}\BibitemShut {NoStop}%
\bibitem [{\citenamefont {Scazza}\ \emph {et~al.}(2017)\citenamefont {Scazza},
  \citenamefont {Valtolina}, \citenamefont {Massignan}, \citenamefont {Recati},
  \citenamefont {Amico}, \citenamefont {Burchianti}, \citenamefont {Fort},
  \citenamefont {Inguscio}, \citenamefont {Zaccanti},\ and\ \citenamefont
  {Roati}}]{Scazza2017PRL}%
  \BibitemOpen
  \bibfield  {author} {\bibinfo {author} {\bibfnamefont {F.}~\bibnamefont
  {Scazza}}, \bibinfo {author} {\bibfnamefont {G.}~\bibnamefont {Valtolina}},
  \bibinfo {author} {\bibfnamefont {P.}~\bibnamefont {Massignan}}, \bibinfo
  {author} {\bibfnamefont {A.}~\bibnamefont {Recati}}, \bibinfo {author}
  {\bibfnamefont {A.}~\bibnamefont {Amico}}, \bibinfo {author} {\bibfnamefont
  {A.}~\bibnamefont {Burchianti}}, \bibinfo {author} {\bibfnamefont
  {C.}~\bibnamefont {Fort}}, \bibinfo {author} {\bibfnamefont {M.}~\bibnamefont
  {Inguscio}}, \bibinfo {author} {\bibfnamefont {M.}~\bibnamefont {Zaccanti}},
  \ and\ \bibinfo {author} {\bibfnamefont {G.}~\bibnamefont {Roati}},\
  }\bibfield  {title} {\enquote {\bibinfo {title} {Repulsive fermi polarons in
  a resonant mixture of ultracold $^{6}\mathrm{Li}$ atoms},}\ }\href@noop {}
  {\bibfield  {journal} {\bibinfo  {journal} {Phys. Rev. Lett.}\ }\textbf
  {\bibinfo {volume} {118}},\ \bibinfo {pages} {083602} (\bibinfo {year}
  {2017})}\BibitemShut {NoStop}%
\bibitem [{\citenamefont {Yan}\ \emph {et~al.}(2019)\citenamefont {Yan},
  \citenamefont {Patel}, \citenamefont {Mukherjee}, \citenamefont {Fletcher},
  \citenamefont {Struck},\ and\ \citenamefont {Zwierlein}}]{Yan2019PRL}%
  \BibitemOpen
  \bibfield  {author} {\bibinfo {author} {\bibfnamefont {Zhenjie}\ \bibnamefont
  {Yan}}, \bibinfo {author} {\bibfnamefont {Parth~B.}\ \bibnamefont {Patel}},
  \bibinfo {author} {\bibfnamefont {Biswaroop}\ \bibnamefont {Mukherjee}},
  \bibinfo {author} {\bibfnamefont {Richard~J.}\ \bibnamefont {Fletcher}},
  \bibinfo {author} {\bibfnamefont {Julian}\ \bibnamefont {Struck}}, \ and\
  \bibinfo {author} {\bibfnamefont {Martin~W.}\ \bibnamefont {Zwierlein}},\
  }\bibfield  {title} {\enquote {\bibinfo {title} {Boiling a unitary fermi
  liquid},}\ }\href@noop {} {\bibfield  {journal} {\bibinfo  {journal} {Phys.
  Rev. Lett.}\ }\textbf {\bibinfo {volume} {122}},\ \bibinfo {pages} {093401}
  (\bibinfo {year} {2019})}\BibitemShut {NoStop}%
\bibitem [{\citenamefont {Yan}\ \emph {et~al.}(2020)\citenamefont {Yan},
  \citenamefont {Ni}, \citenamefont {Robens},\ and\ \citenamefont
  {Zwierlein}}]{Zwierlein2020Science}%
  \BibitemOpen
  \bibfield  {author} {\bibinfo {author} {\bibfnamefont {Z.~Z.}\ \bibnamefont
  {Yan}}, \bibinfo {author} {\bibfnamefont {Y.}~\bibnamefont {Ni}}, \bibinfo
  {author} {\bibfnamefont {C.}~\bibnamefont {Robens}}, \ and\ \bibinfo {author}
  {\bibfnamefont {M.W.}\ \bibnamefont {Zwierlein}},\ }\bibfield  {title}
  {\enquote {\bibinfo {title} {Bose polarons near quantum criticality},}\
  }\href@noop {} {\bibfield  {journal} {\bibinfo  {journal} {Science}\ }\textbf
  {\bibinfo {volume} {368}},\ \bibinfo {pages} {190} (\bibinfo {year}
  {2020})}\BibitemShut {NoStop}%
\bibitem [{\citenamefont {Ness}\ \emph {et~al.}(2020)\citenamefont {Ness},
  \citenamefont {Shkedrov}, \citenamefont {Florshaim}, \citenamefont {Diessel},
  \citenamefont {von Milczewski}, \citenamefont {Schmidt},\ and\ \citenamefont
  {Sagi}}]{Sagi2020PRX}%
  \BibitemOpen
  \bibfield  {author} {\bibinfo {author} {\bibfnamefont {Gal}\ \bibnamefont
  {Ness}}, \bibinfo {author} {\bibfnamefont {Constantine}\ \bibnamefont
  {Shkedrov}}, \bibinfo {author} {\bibfnamefont {Yanay}\ \bibnamefont
  {Florshaim}}, \bibinfo {author} {\bibfnamefont {Oriana~K.}\ \bibnamefont
  {Diessel}}, \bibinfo {author} {\bibfnamefont {Jonas}\ \bibnamefont {von
  Milczewski}}, \bibinfo {author} {\bibfnamefont {Richard}\ \bibnamefont
  {Schmidt}}, \ and\ \bibinfo {author} {\bibfnamefont {Yoav}\ \bibnamefont
  {Sagi}},\ }\bibfield  {title} {\enquote {\bibinfo {title} {Observation of a
  smooth polaron-molecule transition in a degenerate {Fermi} gas},}\
  }\href@noop {} {\bibfield  {journal} {\bibinfo  {journal} {Phys. Rev. X}\
  }\textbf {\bibinfo {volume} {10}},\ \bibinfo {pages} {041019} (\bibinfo
  {year} {2020})}\BibitemShut {NoStop}%
\bibitem [{\citenamefont {Chevy}(2006)}]{Chevy2006PRA}%
  \BibitemOpen
  \bibfield  {author} {\bibinfo {author} {\bibfnamefont {F.}~\bibnamefont
  {Chevy}},\ }\bibfield  {title} {\enquote {\bibinfo {title} {Universal phase
  diagram of a strongly interacting {Fermi} gas with unbalanced spin
  populations},}\ }\href@noop {} {\bibfield  {journal} {\bibinfo  {journal}
  {Phys. Rev. A}\ }\textbf {\bibinfo {volume} {74}},\ \bibinfo {pages} {063628}
  (\bibinfo {year} {2006})}\BibitemShut {NoStop}%
\bibitem [{\citenamefont {Lobo}\ \emph {et~al.}(2006)\citenamefont {Lobo},
  \citenamefont {Recati}, \citenamefont {Giorgini},\ and\ \citenamefont
  {Stringari}}]{Lobo2006PRL}%
  \BibitemOpen
  \bibfield  {author} {\bibinfo {author} {\bibfnamefont {C.}~\bibnamefont
  {Lobo}}, \bibinfo {author} {\bibfnamefont {A.}~\bibnamefont {Recati}},
  \bibinfo {author} {\bibfnamefont {S.}~\bibnamefont {Giorgini}}, \ and\
  \bibinfo {author} {\bibfnamefont {S.}~\bibnamefont {Stringari}},\ }\bibfield
  {title} {\enquote {\bibinfo {title} {Normal state of a polarized {Fermi} gas
  at unitarity},}\ }\href@noop {} {\bibfield  {journal} {\bibinfo  {journal}
  {Phys. Rev. Lett.}\ }\textbf {\bibinfo {volume} {97}},\ \bibinfo {pages}
  {200403} (\bibinfo {year} {2006})}\BibitemShut {NoStop}%
\bibitem [{\citenamefont {Combescot}\ \emph {et~al.}(2007)\citenamefont
  {Combescot}, \citenamefont {Recati}, \citenamefont {Lobo},\ and\
  \citenamefont {Chevy}}]{Combescot2007PRL}%
  \BibitemOpen
  \bibfield  {author} {\bibinfo {author} {\bibfnamefont {R.}~\bibnamefont
  {Combescot}}, \bibinfo {author} {\bibfnamefont {A.}~\bibnamefont {Recati}},
  \bibinfo {author} {\bibfnamefont {C.}~\bibnamefont {Lobo}}, \ and\ \bibinfo
  {author} {\bibfnamefont {F.}~\bibnamefont {Chevy}},\ }\bibfield  {title}
  {\enquote {\bibinfo {title} {Normal state of highly polarized {Fermi} gases:
  Simple many-body approaches},}\ }\href@noop {} {\bibfield  {journal}
  {\bibinfo  {journal} {Phys. Rev. Lett.}\ }\textbf {\bibinfo {volume} {98}},\
  \bibinfo {pages} {180402} (\bibinfo {year} {2007})}\BibitemShut {NoStop}%
\bibitem [{\citenamefont {Punk}\ \emph {et~al.}(2009)\citenamefont {Punk},
  \citenamefont {Dumitrescu},\ and\ \citenamefont {Zwerger}}]{Punk2009PRA}%
  \BibitemOpen
  \bibfield  {author} {\bibinfo {author} {\bibfnamefont {M.}~\bibnamefont
  {Punk}}, \bibinfo {author} {\bibfnamefont {P.~T.}\ \bibnamefont
  {Dumitrescu}}, \ and\ \bibinfo {author} {\bibfnamefont {W.}~\bibnamefont
  {Zwerger}},\ }\bibfield  {title} {\enquote {\bibinfo {title}
  {Polaron-to-molecule transition in a strongly imbalanced {Fermi} gas},}\
  }\href@noop {} {\bibfield  {journal} {\bibinfo  {journal} {Phys. Rev. A}\
  }\textbf {\bibinfo {volume} {80}},\ \bibinfo {pages} {053605} (\bibinfo
  {year} {2009})}\BibitemShut {NoStop}%
\bibitem [{\citenamefont {Cui}\ and\ \citenamefont {Zhai}(2010)}]{Cui2010PRA}%
  \BibitemOpen
  \bibfield  {author} {\bibinfo {author} {\bibfnamefont {Xiaoling}\
  \bibnamefont {Cui}}\ and\ \bibinfo {author} {\bibfnamefont {Hui}\
  \bibnamefont {Zhai}},\ }\bibfield  {title} {\enquote {\bibinfo {title}
  {Stability of a fully magnetized ferromagnetic state in repulsively
  interacting ultracold {Fermi} gases},}\ }\href@noop {} {\bibfield  {journal}
  {\bibinfo  {journal} {Phys. Rev. A}\ }\textbf {\bibinfo {volume} {81}},\
  \bibinfo {pages} {041602} (\bibinfo {year} {2010})}\BibitemShut {NoStop}%
\bibitem [{\citenamefont {Mathy}\ \emph {et~al.}(2011)\citenamefont {Mathy},
  \citenamefont {Parish},\ and\ \citenamefont {Huse}}]{Mathy2011PRL}%
  \BibitemOpen
  \bibfield  {author} {\bibinfo {author} {\bibfnamefont {Charles J.~M.}\
  \bibnamefont {Mathy}}, \bibinfo {author} {\bibfnamefont {Meera~M.}\
  \bibnamefont {Parish}}, \ and\ \bibinfo {author} {\bibfnamefont {David~A.}\
  \bibnamefont {Huse}},\ }\bibfield  {title} {\enquote {\bibinfo {title}
  {Trimers, molecules, and polarons in mass-imbalanced atomic {Fermi} gases},}\
  }\href@noop {} {\bibfield  {journal} {\bibinfo  {journal} {Phys. Rev. Lett.}\
  }\textbf {\bibinfo {volume} {106}},\ \bibinfo {pages} {166404} (\bibinfo
  {year} {2011})}\BibitemShut {NoStop}%
\bibitem [{\citenamefont {Schmidt}\ \emph {et~al.}(2012)\citenamefont
  {Schmidt}, \citenamefont {Enss}, \citenamefont {Pietil\"a},\ and\
  \citenamefont {Demler}}]{Schmidt2012PRA}%
  \BibitemOpen
  \bibfield  {author} {\bibinfo {author} {\bibfnamefont {Richard}\ \bibnamefont
  {Schmidt}}, \bibinfo {author} {\bibfnamefont {Tilman}\ \bibnamefont {Enss}},
  \bibinfo {author} {\bibfnamefont {Ville}\ \bibnamefont {Pietil\"a}}, \ and\
  \bibinfo {author} {\bibfnamefont {Eugene}\ \bibnamefont {Demler}},\
  }\bibfield  {title} {\enquote {\bibinfo {title} {Fermi polarons in two
  dimensions},}\ }\href@noop {} {\bibfield  {journal} {\bibinfo  {journal}
  {Phys. Rev. A}\ }\textbf {\bibinfo {volume} {85}},\ \bibinfo {pages} {021602}
  (\bibinfo {year} {2012})}\BibitemShut {NoStop}%
\bibitem [{\citenamefont {Rath}\ and\ \citenamefont
  {Schmidt}(2013)}]{Rath2013PRA}%
  \BibitemOpen
  \bibfield  {author} {\bibinfo {author} {\bibfnamefont {Steffen~Patrick}\
  \bibnamefont {Rath}}\ and\ \bibinfo {author} {\bibfnamefont {Richard}\
  \bibnamefont {Schmidt}},\ }\bibfield  {title} {\enquote {\bibinfo {title}
  {Field-theoretical study of the {Bose} polaron},}\ }\href@noop {} {\bibfield
  {journal} {\bibinfo  {journal} {Phys. Rev. A}\ }\textbf {\bibinfo {volume}
  {88}},\ \bibinfo {pages} {053632} (\bibinfo {year} {2013})}\BibitemShut
  {NoStop}%
\bibitem [{\citenamefont {Shashi}\ \emph {et~al.}(2014)\citenamefont {Shashi},
  \citenamefont {Grusdt}, \citenamefont {Abanin},\ and\ \citenamefont
  {Demler}}]{Shashi2014PRA}%
  \BibitemOpen
  \bibfield  {author} {\bibinfo {author} {\bibfnamefont {Aditya}\ \bibnamefont
  {Shashi}}, \bibinfo {author} {\bibfnamefont {Fabian}\ \bibnamefont {Grusdt}},
  \bibinfo {author} {\bibfnamefont {Dmitry~A.}\ \bibnamefont {Abanin}}, \ and\
  \bibinfo {author} {\bibfnamefont {Eugene}\ \bibnamefont {Demler}},\
  }\bibfield  {title} {\enquote {\bibinfo {title} {Radio-frequency spectroscopy
  of polarons in ultracold {Bose} gases},}\ }\href@noop {} {\bibfield
  {journal} {\bibinfo  {journal} {Phys. Rev. A}\ }\textbf {\bibinfo {volume}
  {89}},\ \bibinfo {pages} {053617} (\bibinfo {year} {2014})}\BibitemShut
  {NoStop}%
\bibitem [{\citenamefont {Li}\ and\ \citenamefont
  {Das~Sarma}(2014)}]{Li2014PRA}%
  \BibitemOpen
  \bibfield  {author} {\bibinfo {author} {\bibfnamefont {Weiran}\ \bibnamefont
  {Li}}\ and\ \bibinfo {author} {\bibfnamefont {S.}~\bibnamefont {Das~Sarma}},\
  }\bibfield  {title} {\enquote {\bibinfo {title} {Variational study of
  polarons in {Bose-Einstein} condensates},}\ }\href@noop {} {\bibfield
  {journal} {\bibinfo  {journal} {Phys. Rev. A}\ }\textbf {\bibinfo {volume}
  {90}},\ \bibinfo {pages} {013618} (\bibinfo {year} {2014})}\BibitemShut
  {NoStop}%
\bibitem [{\citenamefont {Kroiss}\ and\ \citenamefont
  {Pollet}(2015)}]{Kroiss2015PRL}%
  \BibitemOpen
  \bibfield  {author} {\bibinfo {author} {\bibfnamefont {Peter}\ \bibnamefont
  {Kroiss}}\ and\ \bibinfo {author} {\bibfnamefont {Lode}\ \bibnamefont
  {Pollet}},\ }\bibfield  {title} {\enquote {\bibinfo {title} {Diagrammatic
  monte carlo study of a mass-imbalanced fermi-polaron system},}\ }\href@noop
  {} {\bibfield  {journal} {\bibinfo  {journal} {Phys. Rev. B}\ }\textbf
  {\bibinfo {volume} {91}},\ \bibinfo {pages} {144507} (\bibinfo {year}
  {2015})}\BibitemShut {NoStop}%
\bibitem [{\citenamefont {Levinsen}\ \emph {et~al.}(2015)\citenamefont
  {Levinsen}, \citenamefont {Parish},\ and\ \citenamefont
  {Bruun}}]{Levinsen2015PRL}%
  \BibitemOpen
  \bibfield  {author} {\bibinfo {author} {\bibfnamefont {Jesper}\ \bibnamefont
  {Levinsen}}, \bibinfo {author} {\bibfnamefont {Meera~M.}\ \bibnamefont
  {Parish}}, \ and\ \bibinfo {author} {\bibfnamefont {Georg~M.}\ \bibnamefont
  {Bruun}},\ }\bibfield  {title} {\enquote {\bibinfo {title} {Impurity in a
  bose-einstein condensate and the efimov effect},}\ }\href@noop {} {\bibfield
  {journal} {\bibinfo  {journal} {Phys. Rev. Lett.}\ }\textbf {\bibinfo
  {volume} {115}},\ \bibinfo {pages} {125302} (\bibinfo {year}
  {2015})}\BibitemShut {NoStop}%
\bibitem [{\citenamefont {Hu}\ \emph {et~al.}(2016{\natexlab{b}})\citenamefont
  {Hu}, \citenamefont {Wang}, \citenamefont {Yi},\ and\ \citenamefont
  {Liu}}]{HuHui2016PRA}%
  \BibitemOpen
  \bibfield  {author} {\bibinfo {author} {\bibfnamefont {Hui}\ \bibnamefont
  {Hu}}, \bibinfo {author} {\bibfnamefont {An-Bang}\ \bibnamefont {Wang}},
  \bibinfo {author} {\bibfnamefont {Su}~\bibnamefont {Yi}}, \ and\ \bibinfo
  {author} {\bibfnamefont {Xia-Ji}\ \bibnamefont {Liu}},\ }\bibfield  {title}
  {\enquote {\bibinfo {title} {Fermi polaron in a one-dimensional quasiperiodic
  optical lattice: The simplest many-body localization challenge},}\
  }\href@noop {} {\bibfield  {journal} {\bibinfo  {journal} {Phys. Rev. A}\
  }\textbf {\bibinfo {volume} {93}},\ \bibinfo {pages} {053601} (\bibinfo
  {year} {2016}{\natexlab{b}})}\BibitemShut {NoStop}%
\bibitem [{\citenamefont {Goulko}\ \emph {et~al.}(2016)\citenamefont {Goulko},
  \citenamefont {Mishchenko}, \citenamefont {Prokof'ev},\ and\ \citenamefont
  {Svistunov}}]{Goulko2016PRA}%
  \BibitemOpen
  \bibfield  {author} {\bibinfo {author} {\bibfnamefont {Olga}\ \bibnamefont
  {Goulko}}, \bibinfo {author} {\bibfnamefont {Andrey~S.}\ \bibnamefont
  {Mishchenko}}, \bibinfo {author} {\bibfnamefont {Nikolay}\ \bibnamefont
  {Prokof'ev}}, \ and\ \bibinfo {author} {\bibfnamefont {Boris}\ \bibnamefont
  {Svistunov}},\ }\bibfield  {title} {\enquote {\bibinfo {title} {Dark
  continuum in the spectral function of the resonant fermi polaron},}\
  }\href@noop {} {\bibfield  {journal} {\bibinfo  {journal} {Phys. Rev. A}\
  }\textbf {\bibinfo {volume} {94}},\ \bibinfo {pages} {051605} (\bibinfo
  {year} {2016})}\BibitemShut {NoStop}%
\bibitem [{\citenamefont {Hu}\ \emph {et~al.}(2018)\citenamefont {Hu},
  \citenamefont {Mulkerin}, \citenamefont {Wang},\ and\ \citenamefont
  {Liu}}]{HuHui2018PRA}%
  \BibitemOpen
  \bibfield  {author} {\bibinfo {author} {\bibfnamefont {Hui}\ \bibnamefont
  {Hu}}, \bibinfo {author} {\bibfnamefont {Brendan~C.}\ \bibnamefont
  {Mulkerin}}, \bibinfo {author} {\bibfnamefont {Jia}\ \bibnamefont {Wang}}, \
  and\ \bibinfo {author} {\bibfnamefont {Xia-Ji}\ \bibnamefont {Liu}},\
  }\bibfield  {title} {\enquote {\bibinfo {title} {Attractive fermi polarons at
  nonzero temperatures with a finite impurity concentration},}\ }\href@noop {}
  {\bibfield  {journal} {\bibinfo  {journal} {Phys. Rev. A}\ }\textbf {\bibinfo
  {volume} {98}},\ \bibinfo {pages} {013626} (\bibinfo {year}
  {2018})}\BibitemShut {NoStop}%
\bibitem [{\citenamefont {Pe\~na Ardila}\ \emph {et~al.}(2019)\citenamefont
  {Pe\~na Ardila}, \citenamefont {J\o{}rgensen}, \citenamefont {Pohl},
  \citenamefont {Giorgini}, \citenamefont {Bruun},\ and\ \citenamefont
  {Arlt}}]{PenaArdila2019PRA}%
  \BibitemOpen
  \bibfield  {author} {\bibinfo {author} {\bibfnamefont {L.~A.}\ \bibnamefont
  {Pe\~na Ardila}}, \bibinfo {author} {\bibfnamefont {N.~B.}\ \bibnamefont
  {J\o{}rgensen}}, \bibinfo {author} {\bibfnamefont {T.}~\bibnamefont {Pohl}},
  \bibinfo {author} {\bibfnamefont {S.}~\bibnamefont {Giorgini}}, \bibinfo
  {author} {\bibfnamefont {G.~M.}\ \bibnamefont {Bruun}}, \ and\ \bibinfo
  {author} {\bibfnamefont {J.~J.}\ \bibnamefont {Arlt}},\ }\bibfield  {title}
  {\enquote {\bibinfo {title} {Analyzing a bose polaron across resonant
  interactions},}\ }\href {\doibase 10.1103/PhysRevA.99.063607} {\bibfield
  {journal} {\bibinfo  {journal} {Phys. Rev. A}\ }\textbf {\bibinfo {volume}
  {99}},\ \bibinfo {pages} {063607} (\bibinfo {year} {2019})}\BibitemShut
  {NoStop}%
\bibitem [{\citenamefont {Mulkerin}\ \emph {et~al.}(2019)\citenamefont
  {Mulkerin}, \citenamefont {Liu},\ and\ \citenamefont
  {Hu}}]{Mulkerin2019AnnPhys}%
  \BibitemOpen
  \bibfield  {author} {\bibinfo {author} {\bibfnamefont {B.~C.}\ \bibnamefont
  {Mulkerin}}, \bibinfo {author} {\bibfnamefont {X.-J.}\ \bibnamefont {Liu}}, \
  and\ \bibinfo {author} {\bibfnamefont {H.}~\bibnamefont {Hu}},\ }\bibfield
  {title} {\enquote {\bibinfo {title} {Breakdown of the fermi polaron
  description near fermi degeneracy at unitarity},}\ }\href@noop {} {\bibfield
  {journal} {\bibinfo  {journal} {Ann. Phys. (NY)}\ }\textbf {\bibinfo {volume}
  {407}},\ \bibinfo {pages} {29} (\bibinfo {year} {2019})}\BibitemShut
  {NoStop}%
\bibitem [{\citenamefont {Wang}\ \emph {et~al.}(2019)\citenamefont {Wang},
  \citenamefont {Liu},\ and\ \citenamefont {Hu}}]{Jia2019PRL}%
  \BibitemOpen
  \bibfield  {author} {\bibinfo {author} {\bibfnamefont {Jia}\ \bibnamefont
  {Wang}}, \bibinfo {author} {\bibfnamefont {Xia-Ji}\ \bibnamefont {Liu}}, \
  and\ \bibinfo {author} {\bibfnamefont {Hui}\ \bibnamefont {Hu}},\ }\bibfield
  {title} {\enquote {\bibinfo {title} {Roton-induced bose polaron in the
  presence of synthetic spin-orbit coupling},}\ }\href@noop {} {\bibfield
  {journal} {\bibinfo  {journal} {Phys. Rev. Lett.}\ }\textbf {\bibinfo
  {volume} {123}},\ \bibinfo {pages} {213401} (\bibinfo {year}
  {2019})}\BibitemShut {NoStop}%
\bibitem [{\citenamefont {Isaule}\ \emph {et~al.}(2021)\citenamefont {Isaule},
  \citenamefont {Morera}, \citenamefont {Massignan},\ and\ \citenamefont
  {Juli\'a-D\'{\i}az}}]{Isaule2021PRA}%
  \BibitemOpen
  \bibfield  {author} {\bibinfo {author} {\bibfnamefont {Felipe}\ \bibnamefont
  {Isaule}}, \bibinfo {author} {\bibfnamefont {Ivan}\ \bibnamefont {Morera}},
  \bibinfo {author} {\bibfnamefont {Pietro}\ \bibnamefont {Massignan}}, \ and\
  \bibinfo {author} {\bibfnamefont {Bruno}\ \bibnamefont {Juli\'a-D\'{\i}az}},\
  }\bibfield  {title} {\enquote {\bibinfo {title} {Renormalization-group study
  of bose polarons},}\ }\href@noop {} {\bibfield  {journal} {\bibinfo
  {journal} {Phys. Rev. A}\ }\textbf {\bibinfo {volume} {104}},\ \bibinfo
  {pages} {023317} (\bibinfo {year} {2021})}\BibitemShut {NoStop}%
\bibitem [{\citenamefont {Pessoa}\ \emph {et~al.}(2021)\citenamefont {Pessoa},
  \citenamefont {Vitiello},\ and\ \citenamefont {Ardila}}]{Pessoa2021PRA}%
  \BibitemOpen
  \bibfield  {author} {\bibinfo {author} {\bibfnamefont {Renato}\ \bibnamefont
  {Pessoa}}, \bibinfo {author} {\bibfnamefont {S.~A.}\ \bibnamefont
  {Vitiello}}, \ and\ \bibinfo {author} {\bibfnamefont {L.~A. Pe\~na}\
  \bibnamefont {Ardila}},\ }\bibfield  {title} {\enquote {\bibinfo {title}
  {Finite-range effects in the unitary fermi polaron},}\ }\href {\doibase
  10.1103/PhysRevA.104.043313} {\bibfield  {journal} {\bibinfo  {journal}
  {Phys. Rev. A}\ }\textbf {\bibinfo {volume} {104}},\ \bibinfo {pages}
  {043313} (\bibinfo {year} {2021})}\BibitemShut {NoStop}%
\bibitem [{\citenamefont {Seetharam}\ \emph {et~al.}(2021)\citenamefont
  {Seetharam}, \citenamefont {Shchadilova}, \citenamefont {Grusdt},
  \citenamefont {Zvonarev},\ and\ \citenamefont {Demler}}]{Seetharam2021PRL}%
  \BibitemOpen
  \bibfield  {author} {\bibinfo {author} {\bibfnamefont {Kushal}\ \bibnamefont
  {Seetharam}}, \bibinfo {author} {\bibfnamefont {Yulia}\ \bibnamefont
  {Shchadilova}}, \bibinfo {author} {\bibfnamefont {Fabian}\ \bibnamefont
  {Grusdt}}, \bibinfo {author} {\bibfnamefont {Mikhail~B.}\ \bibnamefont
  {Zvonarev}}, \ and\ \bibinfo {author} {\bibfnamefont {Eugene}\ \bibnamefont
  {Demler}},\ }\bibfield  {title} {\enquote {\bibinfo {title} {Dynamical
  quantum cherenkov transition of fast impurities in quantum liquids},}\ }\href
  {\doibase 10.1103/PhysRevLett.127.185302} {\bibfield  {journal} {\bibinfo
  {journal} {Phys. Rev. Lett.}\ }\textbf {\bibinfo {volume} {127}},\ \bibinfo
  {pages} {185302} (\bibinfo {year} {2021})}\BibitemShut {NoStop}%
\bibitem [{\citenamefont {Nishida}(2015)}]{Nishida2015PRL}%
  \BibitemOpen
  \bibfield  {author} {\bibinfo {author} {\bibfnamefont {Yusuke}\ \bibnamefont
  {Nishida}},\ }\bibfield  {title} {\enquote {\bibinfo {title} {Polaronic
  atom-trimer continuity in three-component fermi gases},}\ }\href@noop {}
  {\bibfield  {journal} {\bibinfo  {journal} {Phys. Rev. Lett.}\ }\textbf
  {\bibinfo {volume} {114}},\ \bibinfo {pages} {115302} (\bibinfo {year}
  {2015})}\BibitemShut {NoStop}%
\bibitem [{\citenamefont {Yi}\ and\ \citenamefont {Cui}(2015)}]{Yi2015PRA}%
  \BibitemOpen
  \bibfield  {author} {\bibinfo {author} {\bibfnamefont {Wei}\ \bibnamefont
  {Yi}}\ and\ \bibinfo {author} {\bibfnamefont {Xiaoling}\ \bibnamefont
  {Cui}},\ }\bibfield  {title} {\enquote {\bibinfo {title} {Polarons in
  ultracold fermi superfluids},}\ }\href@noop {} {\bibfield  {journal}
  {\bibinfo  {journal} {Phys. Rev. A}\ }\textbf {\bibinfo {volume} {92}},\
  \bibinfo {pages} {013620} (\bibinfo {year} {2015})}\BibitemShut {NoStop}%
\bibitem [{\citenamefont {Pierce}\ \emph {et~al.}(2019)\citenamefont {Pierce},
  \citenamefont {Leyronas},\ and\ \citenamefont {Chevy}}]{Pierce2019PRL}%
  \BibitemOpen
  \bibfield  {author} {\bibinfo {author} {\bibfnamefont {M.}~\bibnamefont
  {Pierce}}, \bibinfo {author} {\bibfnamefont {X.}~\bibnamefont {Leyronas}}, \
  and\ \bibinfo {author} {\bibfnamefont {F.}~\bibnamefont {Chevy}},\ }\bibfield
   {title} {\enquote {\bibinfo {title} {Few versus many-body physics of an
  impurity immersed in a superfluid of spin $1/2$ attractive fermions},}\
  }\href {\doibase 10.1103/PhysRevLett.123.080403} {\bibfield  {journal}
  {\bibinfo  {journal} {Phys. Rev. Lett.}\ }\textbf {\bibinfo {volume} {123}},\
  \bibinfo {pages} {080403} (\bibinfo {year} {2019})}\BibitemShut {NoStop}%
\bibitem [{\citenamefont {Hu}\ \emph {et~al.}(2021)\citenamefont {Hu},
  \citenamefont {Wang}, \citenamefont {Zhou},\ and\ \citenamefont
  {Liu}}]{HuHui2021arXiv}%
  \BibitemOpen
  \bibfield  {author} {\bibinfo {author} {\bibfnamefont {Hui}\ \bibnamefont
  {Hu}}, \bibinfo {author} {\bibfnamefont {Jia}\ \bibnamefont {Wang}}, \bibinfo
  {author} {\bibfnamefont {Jing}\ \bibnamefont {Zhou}}, \ and\ \bibinfo
  {author} {\bibfnamefont {Xia-Ji}\ \bibnamefont {Liu}},\ }\href@noop {}
  {\enquote {\bibinfo {title} {Crossover polarons in a strongly interacting
  fermi superfluid},}\ } (\bibinfo {year} {2021}),\ \bibinfo {note} {arXiv:
  2111.01372}\BibitemShut {NoStop}%
\bibitem [{\citenamefont {Bigu\'{e}}\ \emph {et~al.}(2022)\citenamefont
  {Bigu\'{e}}, \citenamefont {Chevy},\ and\ \citenamefont
  {Leyronas}}]{Bigue2022arXiv}%
  \BibitemOpen
  \bibfield  {author} {\bibinfo {author} {\bibfnamefont {A.}~\bibnamefont
  {Bigu\'{e}}}, \bibinfo {author} {\bibfnamefont {F.}~\bibnamefont {Chevy}}, \
  and\ \bibinfo {author} {\bibfnamefont {X.}~\bibnamefont {Leyronas}},\
  }\href@noop {} {\enquote {\bibinfo {title} {Mean-field vs rpa calculation of
  the energy of an impurity immersed in a spin 1/2 superfluid},}\ } (\bibinfo
  {year} {2022}),\ \bibinfo {note} {arXiv: 2202.03222}\BibitemShut {NoStop}%
\bibitem [{\citenamefont {Nozi{\`e}res}\ and\ \citenamefont {{De
  Dominics}}(1969)}]{Nozieres1969PR}%
  \BibitemOpen
  \bibfield  {author} {\bibinfo {author} {\bibfnamefont {P.}~\bibnamefont
  {Nozi{\`e}res}}\ and\ \bibinfo {author} {\bibfnamefont {C.~T.}\ \bibnamefont
  {{De Dominics}}},\ }\bibfield  {title} {\enquote {\bibinfo {title}
  {Singularities in the x-ray absorption and emission of metals. iii. one-body
  theory exact solution},}\ }\href@noop {} {\bibfield  {journal} {\bibinfo
  {journal} {Phys. Rev.}\ }\textbf {\bibinfo {volume} {178}},\ \bibinfo {pages}
  {1097--1107} (\bibinfo {year} {1969})}\BibitemShut {NoStop}%
\bibitem [{\citenamefont {Mahan}(2000)}]{Mahan2000Book}%
  \BibitemOpen
  \bibfield  {author} {\bibinfo {author} {\bibfnamefont {Gerald~D.}\
  \bibnamefont {Mahan}},\ }\href@noop {} {\emph {\bibinfo {title} {Many
  Particle Physics}}},\ \bibinfo {edition} {3rd}\ ed.\ (\bibinfo  {publisher}
  {Kluwer},\ \bibinfo {address} {New York},\ \bibinfo {year}
  {2000})\BibitemShut {NoStop}%
\bibitem [{\citenamefont {Weiss}(1999)}]{Weiss1999Book}%
  \BibitemOpen
  \bibfield  {author} {\bibinfo {author} {\bibfnamefont {U.}~\bibnamefont
  {Weiss}},\ }\href@noop {} {\emph {\bibinfo {title} {Quantum Dissipative
  Systems}}},\ Vol.~\bibinfo {volume} {10}\ (\bibinfo  {publisher} {World
  Scientific},\ \bibinfo {address} {Singapore},\ \bibinfo {year}
  {1999})\BibitemShut {NoStop}%
\bibitem [{\citenamefont {Rosch}(1999)}]{Rosch1999AdvPhys}%
  \BibitemOpen
  \bibfield  {author} {\bibinfo {author} {\bibfnamefont {A.}~\bibnamefont
  {Rosch}},\ }\bibfield  {title} {\enquote {\bibinfo {title} {Quantum-coherent
  transport of a heavy particle in a fermionic bath},}\ }\href@noop {}
  {\bibfield  {journal} {\bibinfo  {journal} {Adv. Phys.}\ }\textbf {\bibinfo
  {volume} {48}},\ \bibinfo {pages} {295} (\bibinfo {year} {1999})}\BibitemShut
  {NoStop}%
\bibitem [{\citenamefont {Anderson}(1967)}]{Anderson1967PRL}%
  \BibitemOpen
  \bibfield  {author} {\bibinfo {author} {\bibfnamefont {P.~W.}\ \bibnamefont
  {Anderson}},\ }\bibfield  {title} {\enquote {\bibinfo {title} {Infrared
  catastrophe in fermi gases with local scattering potentials},}\ }\href@noop
  {} {\bibfield  {journal} {\bibinfo  {journal} {Phys. Rev. Lett.}\ }\textbf
  {\bibinfo {volume} {18}},\ \bibinfo {pages} {1049--1051} (\bibinfo {year}
  {1967})}\BibitemShut {NoStop}%
\bibitem [{\citenamefont {Levitov}\ and\ \citenamefont
  {Lee}(1996)}]{Leonid1996JMathPhys}%
  \BibitemOpen
  \bibfield  {author} {\bibinfo {author} {\bibfnamefont {Leonid~S.}\
  \bibnamefont {Levitov}}\ and\ \bibinfo {author} {\bibfnamefont {Hyunwoo}\
  \bibnamefont {Lee}},\ }\bibfield  {title} {\enquote {\bibinfo {title}
  {Electron counting statistics and coherent states of electric current},}\
  }\href@noop {} {\bibfield  {journal} {\bibinfo  {journal} {J. Math. Phys.}\
  }\textbf {\bibinfo {volume} {37}},\ \bibinfo {pages} {4845} (\bibinfo {year}
  {1996})}\BibitemShut {NoStop}%
\bibitem [{\citenamefont {Klich}(2003)}]{Klich2003Book}%
  \BibitemOpen
  \bibfield  {author} {\bibinfo {author} {\bibfnamefont {I.}~\bibnamefont
  {Klich}},\ }\href@noop {} {\emph {\bibinfo {title} {Full Counting Statistics:
  an Elementary Derivation of {Levitov's} Formula}}}\ (\bibinfo  {publisher}
  {Kluwer},\ \bibinfo {address} {Dordrecht},\ \bibinfo {year}
  {2003})\BibitemShut {NoStop}%
\bibitem [{\citenamefont {Sch{\"o}nhammer}(2007)}]{Schonhammer2007PRB}%
  \BibitemOpen
  \bibfield  {author} {\bibinfo {author} {\bibfnamefont {K.}~\bibnamefont
  {Sch{\"o}nhammer}},\ }\bibfield  {title} {\enquote {\bibinfo {title} {Full
  counting statistics for noninteracting fermions: Exact results and the
  {Levitov-Lesovik} formula},}\ }\href@noop {} {\bibfield  {journal} {\bibinfo
  {journal} {Phys. Rev. B}\ }\textbf {\bibinfo {volume} {75}},\ \bibinfo
  {pages} {205329} (\bibinfo {year} {2007})}\BibitemShut {NoStop}%
\bibitem [{\citenamefont {Ivanov}\ and\ \citenamefont
  {Abanov}(2013)}]{Ivanov2013JMathPhys}%
  \BibitemOpen
  \bibfield  {author} {\bibinfo {author} {\bibfnamefont {Dmitri~A}\
  \bibnamefont {Ivanov}}\ and\ \bibinfo {author} {\bibfnamefont {Alexander~G}\
  \bibnamefont {Abanov}},\ }\bibfield  {title} {\enquote {\bibinfo {title}
  {{Fisher-Hartwig} expansion for {Toeplitz} determinants and the spectrum of a
  single-particle reduced density matrix for one-dimensional free fermions},}\
  }\href@noop {} {\bibfield  {journal} {\bibinfo  {journal} {J. Phys. A: Math.
  Theor.}\ }\textbf {\bibinfo {volume} {46}},\ \bibinfo {pages} {375005}
  (\bibinfo {year} {2013})}\BibitemShut {NoStop}%
\bibitem [{\citenamefont {Leggett}(1980)}]{Leggett1980JPhys}%
  \BibitemOpen
  \bibfield  {author} {\bibinfo {author} {\bibfnamefont {A.~J.}\ \bibnamefont
  {Leggett}},\ }\bibfield  {title} {\enquote {\bibinfo {title} {Cooper pairing
  in spin-polarized {Fermi} systems},}\ }\href@noop {} {\bibfield  {journal}
  {\bibinfo  {journal} {J. Phys. (Paris)}\ }\textbf {\bibinfo {volume} {42}},\
  \bibinfo {pages} {C7} (\bibinfo {year} {1980})}\BibitemShut {NoStop}%
\bibitem [{\citenamefont {Nozi{\`e}res}\ and\ \citenamefont
  {Schmitt-Rink}(1985)}]{Nozieres1985JLowTempPhys}%
  \BibitemOpen
  \bibfield  {author} {\bibinfo {author} {\bibfnamefont {P.}~\bibnamefont
  {Nozi{\`e}res}}\ and\ \bibinfo {author} {\bibfnamefont {S.}~\bibnamefont
  {Schmitt-Rink}},\ }\bibfield  {title} {\enquote {\bibinfo {title} {{Bose}
  condensationin an attractive fermion gas: From weak to strong coupling
  superconductivity},}\ }\href@noop {} {\bibfield  {journal} {\bibinfo
  {journal} {J. Low Temp. Phys.}\ }\textbf {\bibinfo {volume} {59}},\ \bibinfo
  {pages} {195} (\bibinfo {year} {1985})}\BibitemShut {NoStop}%
\bibitem [{\citenamefont {Hu}\ \emph {et~al.}(2006)\citenamefont {Hu},
  \citenamefont {Liu},\ and\ \citenamefont {Drummond}}]{HuHui2006EurphysLett}%
  \BibitemOpen
  \bibfield  {author} {\bibinfo {author} {\bibfnamefont {H.}~\bibnamefont
  {Hu}}, \bibinfo {author} {\bibfnamefont {X.-J.}\ \bibnamefont {Liu}}, \ and\
  \bibinfo {author} {\bibfnamefont {P.~D.}\ \bibnamefont {Drummond}},\
  }\bibfield  {title} {\enquote {\bibinfo {title} {Equation of state of a
  superfluid fermi gas in the bcs-bec crossover},}\ }\href@noop {} {\bibfield
  {journal} {\bibinfo  {journal} {Europhys. Lett.}\ }\textbf {\bibinfo {volume}
  {74}},\ \bibinfo {pages} {574} (\bibinfo {year} {2006})}\BibitemShut
  {NoStop}%
\bibitem [{\citenamefont {Yu}(1965)}]{Yu1965ActaPhysSin}%
  \BibitemOpen
  \bibfield  {author} {\bibinfo {author} {\bibfnamefont {L.}~\bibnamefont
  {Yu}},\ }\bibfield  {title} {\enquote {\bibinfo {title} {Bound state in
  superconductors with paramagnetic impurities},}\ }\href@noop {} {\bibfield
  {journal} {\bibinfo  {journal} {Acta. Phys. Sin.}\ }\textbf {\bibinfo
  {volume} {21}},\ \bibinfo {pages} {75} (\bibinfo {year} {1965})}\BibitemShut
  {NoStop}%
\bibitem [{\citenamefont {Shiba}(1968)}]{Shiba1968ProgTheorPhys}%
  \BibitemOpen
  \bibfield  {author} {\bibinfo {author} {\bibfnamefont {H.}~\bibnamefont
  {Shiba}},\ }\bibfield  {title} {\enquote {\bibinfo {title} {Classical spin in
  superconductors},}\ }\href@noop {} {\bibfield  {journal} {\bibinfo  {journal}
  {Prog. Theor. Phys.}\ }\textbf {\bibinfo {volume} {40}},\ \bibinfo {pages}
  {435} (\bibinfo {year} {1968})}\BibitemShut {NoStop}%
\bibitem [{\citenamefont {Rusinov}(1969)}]{Rusinov1969JETP}%
  \BibitemOpen
  \bibfield  {author} {\bibinfo {author} {\bibfnamefont {A.~I.}\ \bibnamefont
  {Rusinov}},\ }\bibfield  {title} {\enquote {\bibinfo {title}
  {Superconductivity near a paramagnetic impurity},}\ }\href@noop {} {\bibfield
   {journal} {\bibinfo  {journal} {JETP Lett. (USSR)}\ }\textbf {\bibinfo
  {volume} {9}},\ \bibinfo {pages} {85} (\bibinfo {year} {1969})}\BibitemShut
  {NoStop}%
\bibitem [{\citenamefont {Vernier}\ \emph {et~al.}(2011)\citenamefont
  {Vernier}, \citenamefont {Pekker}, \citenamefont {Zwierlein},\ and\
  \citenamefont {Demler}}]{Vernier2011PRA}%
  \BibitemOpen
  \bibfield  {author} {\bibinfo {author} {\bibfnamefont {Eric}\ \bibnamefont
  {Vernier}}, \bibinfo {author} {\bibfnamefont {David}\ \bibnamefont {Pekker}},
  \bibinfo {author} {\bibfnamefont {Martin~W.}\ \bibnamefont {Zwierlein}}, \
  and\ \bibinfo {author} {\bibfnamefont {Eugene}\ \bibnamefont {Demler}},\
  }\bibfield  {title} {\enquote {\bibinfo {title} {Bound states of a localized
  magnetic impurity in a superfluid of paired ultracold fermions},}\
  }\href@noop {} {\bibfield  {journal} {\bibinfo  {journal} {Phys. Rev. A}\
  }\textbf {\bibinfo {volume} {83}},\ \bibinfo {pages} {033619} (\bibinfo
  {year} {2011})}\BibitemShut {NoStop}%
\bibitem [{\citenamefont {Jiang}\ \emph {et~al.}(2011)\citenamefont {Jiang},
  \citenamefont {Baksmaty}, \citenamefont {Hu}, \citenamefont {Chen},\ and\
  \citenamefont {Pu}}]{Jiang2011PRA}%
  \BibitemOpen
  \bibfield  {author} {\bibinfo {author} {\bibfnamefont {Lei}\ \bibnamefont
  {Jiang}}, \bibinfo {author} {\bibfnamefont {Leslie~O.}\ \bibnamefont
  {Baksmaty}}, \bibinfo {author} {\bibfnamefont {Hui}\ \bibnamefont {Hu}},
  \bibinfo {author} {\bibfnamefont {Yan}\ \bibnamefont {Chen}}, \ and\ \bibinfo
  {author} {\bibfnamefont {Han}\ \bibnamefont {Pu}},\ }\bibfield  {title}
  {\enquote {\bibinfo {title} {Single impurity in ultracold fermi
  superfluids},}\ }\href {\doibase 10.1103/PhysRevA.83.061604} {\bibfield
  {journal} {\bibinfo  {journal} {Phys. Rev. A}\ }\textbf {\bibinfo {volume}
  {83}},\ \bibinfo {pages} {061604} (\bibinfo {year} {2011})}\BibitemShut
  {NoStop}%
\bibitem [{\citenamefont {Goold}\ \emph {et~al.}(2011)\citenamefont {Goold},
  \citenamefont {Fogarty}, \citenamefont {Lo~Gullo}, \citenamefont
  {Paternostro},\ and\ \citenamefont {Busch}}]{Goold2011PRA}%
  \BibitemOpen
  \bibfield  {author} {\bibinfo {author} {\bibfnamefont {J.}~\bibnamefont
  {Goold}}, \bibinfo {author} {\bibfnamefont {T.}~\bibnamefont {Fogarty}},
  \bibinfo {author} {\bibfnamefont {N.}~\bibnamefont {Lo~Gullo}}, \bibinfo
  {author} {\bibfnamefont {M.}~\bibnamefont {Paternostro}}, \ and\ \bibinfo
  {author} {\bibfnamefont {Th.}\ \bibnamefont {Busch}},\ }\bibfield  {title}
  {\enquote {\bibinfo {title} {Orthogonality catastrophe as a consequence of
  qubit embedding in an ultracold fermi gas},}\ }\href@noop {} {\bibfield
  {journal} {\bibinfo  {journal} {Phys. Rev. A}\ }\textbf {\bibinfo {volume}
  {84}},\ \bibinfo {pages} {063632} (\bibinfo {year} {2011})}\BibitemShut
  {NoStop}%
\bibitem [{\citenamefont {Wang}\ \emph {et~al.}(2022)\citenamefont {Wang},
  \citenamefont {Liu},\ and\ \citenamefont {Hu}}]{AccompanyingLong2022PRA}%
  \BibitemOpen
  \bibfield  {author} {\bibinfo {author} {\bibfnamefont {Jia}\ \bibnamefont
  {Wang}}, \bibinfo {author} {\bibfnamefont {Xia-Ji}\ \bibnamefont {Liu}}, \
  and\ \bibinfo {author} {\bibfnamefont {Hui}\ \bibnamefont {Hu}},\ }\bibfield
  {title} {\enquote {\bibinfo {title} {Heavy polarons in ultracold atomic fermi
  superfluids at the bec-bcs crossover: formalism and applications},}\
  }\href@noop {} {\bibfield  {journal} {\bibinfo  {journal} {submitted to Phys.
  Rev. A.}\ } (\bibinfo {year} {2022})}\BibitemShut {NoStop}%
\bibitem [{Sca()}]{ScatteringLength}%
  \BibitemOpen
  \href@noop {} {}\bibinfo {note} {One should not confuse $a_{\sigma}$ with the
  scattering length at zero scattering energy, although the differences are
  negligible except very close to resonances.}\BibitemShut {Stop}%
\bibitem [{\citenamefont {Ma}(1985)}]{Ma1985PRB}%
  \BibitemOpen
  \bibfield  {author} {\bibinfo {author} {\bibfnamefont {Yanjun}\ \bibnamefont
  {Ma}},\ }\bibfield  {title} {\enquote {\bibinfo {title} {X-ray edges of
  superconducting metals},}\ }\href {\doibase 10.1103/PhysRevB.32.1472}
  {\bibfield  {journal} {\bibinfo  {journal} {Phys. Rev. B}\ }\textbf {\bibinfo
  {volume} {32}},\ \bibinfo {pages} {1472--1475} (\bibinfo {year}
  {1985})}\BibitemShut {NoStop}%
\bibitem [{\citenamefont {Tung}\ \emph {et~al.}(2013)\citenamefont {Tung},
  \citenamefont {Parker}, \citenamefont {Johansen}, \citenamefont {Chin},
  \citenamefont {Wang},\ and\ \citenamefont {Julienne}}]{Tung2013PRA}%
  \BibitemOpen
  \bibfield  {author} {\bibinfo {author} {\bibfnamefont {Shih-Kuang}\
  \bibnamefont {Tung}}, \bibinfo {author} {\bibfnamefont {Colin}\ \bibnamefont
  {Parker}}, \bibinfo {author} {\bibfnamefont {Jacob}\ \bibnamefont
  {Johansen}}, \bibinfo {author} {\bibfnamefont {Cheng}\ \bibnamefont {Chin}},
  \bibinfo {author} {\bibfnamefont {Yujun}\ \bibnamefont {Wang}}, \ and\
  \bibinfo {author} {\bibfnamefont {Paul~S.}\ \bibnamefont {Julienne}},\
  }\bibfield  {title} {\enquote {\bibinfo {title} {Ultracold mixtures of atomic
  ${}^{6}$li and ${}^{133}$cs with tunable interactions},}\ }\href {\doibase
  10.1103/PhysRevA.87.010702} {\bibfield  {journal} {\bibinfo  {journal} {Phys.
  Rev. A}\ }\textbf {\bibinfo {volume} {87}},\ \bibinfo {pages} {010702}
  (\bibinfo {year} {2013})}\BibitemShut {NoStop}%
\end{thebibliography}%

\end{document}